\documentclass[preprint]{aastex63}
\usepackage{amsmath}
\usepackage{amssymb}
\usepackage{graphicx}
\usepackage{color}
\usepackage{xcolor}
\usepackage{xfrac}
\usepackage[caption=false]{subfig}
\usepackage{rotating}
\usepackage{hyperref}
\usepackage[utf8]{inputenc}

\usepackage{etoolbox}
\makeatletter
\patchcmd\linenumberpar{\@LN@parpgbrk}{\penalty\@LN@parpgpen\relax}{}{}
\makeatother

\definecolor{niceblue}{rgb}{0.388235, 0.627451, 0.847059}
\definecolor{nicered}{rgb}{0.7,0.1,0.1}
\definecolor{nicegreen}{rgb}{0.1,0.5,0.1}

\renewcommand{\theequation}{\thesection.\arabic{equation}}

\begin{document}

\title{Two-point correlation function studies 
for the Milky Way: \\
discovery of 
spatial clustering 
from disk excitations and substructure 
} 

\author[0000-0002-9785-914X]{Austin Hinkel}
\affiliation{Department of Physics and Astronomy, University of Kentucky, Lexington, KY 40506}
\affiliation{Department of Physics, 
Colorado College, Colorado Springs, CO 80903}
\author[0000-0002-6166-5546]{Susan Gardner}
\affiliation{Department of Physics and Astronomy, 
University of Kentucky, Lexington, KY 40506}
\author[0000-0002-9541-2678]{Brian Yanny}
\affiliation{Fermi National Accelerator Laboratory, Batavia, IL 60510}

\date{\today}

\begin{abstract}
   We introduce a two-particle correlation function (2PCF) 
   for the Milky Way, 
   constructed to probe spatial correlations in the orthogonal 
   directions of 
   the stellar disk in the Galactic cylindrical coordinates of $R$, $\phi$, and $z$.
   We use this new tool to probe the structure and dynamics of 
   the Galaxy 
   using the carefully selected 
   set of solar neighborhood stars ($d \lesssim 3\, \rm kpc$)
   from Gaia Data Release 2 that we previously 
   employed for studies of axial 
   symmetry breaking in stellar number counts. 
   We make additional, extensive tests, comparing to reference numerical simulations,
   to ensure our control over 
   possibly confounding systematic effects. 
   Supposing either 
   axial or north-south symmetry we divide this data set 
   into two nominally symmetric sectors and construct the 2PCF, 
   in the manner of the Landy--Szalay estimator, from the Gaia data.
   In so doing, working well away from the mid-plane region in which the spiral arms
   appear, we 
   have discovered 
   distinct symmetry-breaking patterns in the 2PCF in its 
   orthogonal directions, 
      thus establishing the 
   existence of correlations in stellar number counts alone
   at subkiloparsec length scales
   for the very first time. 
   In particular, we observe
   extensive 
   wavelike structures of amplitude 
   greatly in excess of what we would estimate if the system were in a steady state. 
   We study the variations in these patterns
   across the Galactic disk, and with increasing $|z|$, 
   and we show  
   how our results complement other observations of non-steady-state effects
   near the Sun, such as vertical asymmetries in stellar number counts and the Gaia snail. 
\end{abstract}

\section{Introduction} \label{sec:Intro}

The long-range nature of the gravitational force distinguishes
the statistical description of stars in the Galaxy from the
terrestrial systems commonly studied. Indeed the 
gravitational ``charge'' cannot be shielded, so that the 
stars 
accelerate, smoothly, through the force field 
dominated by the Galaxy's 
most distant stars. Thus they can be modeled as a 
collisionless fluid with distribution
functions that are assumed to be separable, making 
the stars uncorrelated, even if correlations
can be expected to exist~\citep{binney2008GD}.
For example, stars can be 
formed in spatially and temporally correlated 
ways, as discussed in the 
case of the ``solar family''~\citep{bland-hawthorn2004solarfamily,Bland-Hawthorn2010long-termevolution}, where we note the 
review of \citet{krumholz2019starclusterstime}. 
It is thought, though, that 
these correlations cannot not survive for long in 
the Galactic 
environment~\citep{ladalada2003embeddedclusters,Gieles2006starclusterdisrupt}. 
However, with the 
accumulating evidence for 
apparent wave-like or perturbed features 
in the Galactic disk~\citep{widrow2012galactoseismology,williams2013wobblyRAVE,yanny2013stellar,ferguson2017milky,bennett2018vertical,antoja2018dynamically}, and 
with 
the observed pattern of axial- and 
vertical-symmetry-breaking in the stellar number counts speaking to  
their origin in 
non-steady-state effects~\citep{GHY20,HGY20}, we believe 
the time is now ripe for the development 
of sensitive probes of the structure and dynamics of our Galaxy 
in the region within a few kpc of the Sun. 
In this paper, we 
show how the 
two-point correlation function (2PCF) can be computed in a fully data-driven way 
using symmetry-breaking effects in the stellar number counts and then proceed to make its 
direct determination 
using {\it Gaia} Data Release 2 (DR2) data~\citep{prusti2016gaia2,lindegren2018gaia2}.
We do not employ {\it Gaia} Data Release 3 (DR3) 
because 
we estimate that the improvement in the average precision of the parallax measurements for 
our particular subset of DR2 stars \citep{HGY20} to be less than a few percent, so that 
we think an update is not warranted, 
even if other regions of the Galaxy see marked improvements in parallax precision and 
measurements of 3-D velocity data~\citep{GaiaDR3ref}. We refer to Sec.~\ref{sec:Data} for 
an overview of the features of our data set.  We also note \citet{nelson2022gaussian} for a study of local stellar velocity correlations within a Gaussian process model.

The 2PCF, or the pair correlation function, 
has been used broadly and extensively as a sensitive probe of 
the structure of matter, with applications, e.g., to terrestrial studies in 
condensed matter~\citep{Goodstein2014States} and nuclear~\citep{blatt_weisskopf2010theoretical} physics
and to cosmology~\citep{Peebles1980book}. In the last case it has been used as a 
probe of the large-scale structure of the Universe, through studies of galaxy-galaxy clustering. 
The supposed isotropic nature of the cosmos implies that the 2PCF in this case appears in the terms of
the scalar separation $d$ of any two galaxies in three-dimensional space, 
with the 2PCF capturing the likelihood, given
one galaxy, that another will be found a distance $d$ 
away~\citep{Peebles1973ApJ...185..413P,Hauser1973ApJ...185..757H,Peebles1993ppc..book.....P}. 

In order to develop a suitable 2PCF for studies in our Galaxy, additional considerations enter:  
certainly the Galaxy is not isotropic, and the 2PCF, studied here in its stars, 
can depend on the independent components of the displacement vector between any two of them. 
We develop 
this line of thinking in this paper, starting in Sec.~\ref{sec:Theory}. 
We note that studies of the 2PCF in the Galaxy do exist~\citep{Cooper2011MNRAS_2PCFhalo,Mao2015arXiv150701593M,lancaster2019quantifying,kamdar2021preliminaryclusteringindisk}, though they are 
focused on somewhat different questions and are also limited in different ways. 
An assumption of spherical symmetry has carried over to those studies. In \citet{Cooper2011MNRAS_2PCFhalo} and 
\citet{lancaster2019quantifying} the 2PCF is 
used to study substructure in the stellar halo of the 
Milky Way, with the later, larger study made 
as a probe of the Galaxy's accretion history. There, 
observations of RR Lyrae stars from the CRTS~\citep{Drake2009CatalinaCRTS} survey, 
with 31,301 objects over Galactocentric radii from $\approx 2$ to 90 kpc, and 
from the PanSTARRS1~\citep{Flewelling2016panstarrs1,Chambers2016panstarrs} survey, 
with 44,208 objects over Galactocentric radii from $\approx 0.5$ to 150 kpc, 
are used to determine the 2PCF via the Landy-Szalay (LS)~\citep{landy1993bias} 
correlator. 
The appearance of substructure is inferred through comparison
of the data to a reference 
theoretical background distribution. Extensive study has shown
the LS method to be a superior 
choice~\citep{lancaster2019quantifying,Keihanen2019galaxy_2PCF_splitrandom}, that we and \citet{kamdar2021preliminaryclusteringindisk} also employ, though we note 
\citet{wall2012practical} for discussion of the broader possibilities. In 
\citet{kamdar2021preliminaryclusteringindisk} spatial 
and kinematic clustering of the stars in the Galactic disk 
is studied using a sample of 
$1.7\times 10^6$ stars with 6D phase space information 
within 1 kpc of 
the Sun from {\it Gaia} DR2 data. The construction of a suitable reference distribution 
is essential to the determination of the 2PCF, and in the 
Galaxy its construction is challenging~\citep{Mao2015arXiv150701593M,kamdar2021preliminaryclusteringindisk}. \citet{kamdar2021preliminaryclusteringindisk}
employ Dirichlet process Gaussian mixture models to that end. With that in place, 
they find evidence of clustering of approximately co-moving stars up to large spatial and
kinematic scales, i.e., up to 300 pc and 15 km s$^{-1}$. These results agree well at 
small scales with 
their 
simulations built to model star formation~\citep{Kamdar2019amodelclusteredstar}, and 
their co-natal correlations~\citep{Kamdar2019bmovetogetherborntogether}, in the Galaxy.

Although the relatively bright radial velocity data set employed by \citet{kamdar2021preliminaryclusteringindisk} is  
appropriate to 
their study of young stars close to the Galactic mid-plane, the stars in {\it Gaia} DR2 (and DR3) 
with radial velocity information 
within the Galaxy have sampling biases that could impact the outcomes of our study \citep{katz2019gaia, katz2022gaia}, 
and thus a larger-scale 2PCF analysis requires 
a different approach.
Instead,
we exploit the exceptionally ($>99 \%$) complete {\it Gaia} DR2 sample of \citet{HGY20} 
to effect a data-driven 2PCF analysis of up to some 11.7 million stars, employing spatial 
information only. 
Here we split that data set into sub-samples that are related by symmetry, 
either axial, i.e., in the plane of the Galactic disk, 
or North/South, for our 2PCF analysis, allowing us to focus on the 
spatial correlations in our sample consequent to the existence of spatial 
symmetry breaking effects. For reference, we note that 
the discovery of a vertical wave-like asymmetry 
in the stellar number counts using Sloan Digital Sky Survey (SDSS)~\citep{york2000sdss} data~\citep{widrow2012galactoseismology,yanny2013stellar}, and in the vertical 
velocity distribution from RAVE data~\citep{williams2013wobblyRAVE}, 
with the observed effect changing in different regions  
of the Galactic disk~\citep{ferguson2017milky}, 
hints to a complex vertical landscape in the local Galactic potential. 
Moreover, in 
addition to this picture of planar vertical waves in the Milky Way, \citet{antoja2018dynamically} have shown that a ``phase-space spiral'' pattern in $z-v_{z}$ phase space exists,  
with its very visibility suggesting it is 
a fairly recent development in the Galaxy's past~\citep{antoja2018dynamically}.  
\citet{antoja2018dynamically} 
interpret this snail-shaped pattern using a heuristic, 
anharmonic oscillator model in order to derive an approximate date for the perturbation which is thought to have caused it. 
They find a time-scale of approximately 300-900 Myr, which appears to be consistent with the Sagittarius Dwarf's last passage through the disk \citep{purcell2011sagittarius}, which 
may be a driver of the vertical asymmetries seen \citep{widrow2012galactoseismology,gomez2012vertical}. 
We note, moreover, that 
corrugations across the disk have 
been observed \citep{xu2015rings, blandhawthorn2021snail}. 
Such effects 
may be due to the last impact of the Sagittarius Dwarf 
Galaxy, modulated by the 
influence from the Large Magellanic Cloud (LMC) \citep{laporte2018influence}. 
The Galaxy also has a warp, both in 
its gas~\citep{kerr1957LargescaleMW,burke1957systematic,levine2006vertical} and in its 
stars~\citep{Freudenreich1994DIRBE,drimmel2001three,poggio2018galactic}.
The LMC may also be warping the disk \citep{kerr1957magellanic,weinberg2006magellanic,GHY20}. 
Corrugations of a similar nature are observed in Milky-Way-like galaxies \citep{gomez2021tidally}, and those in the Milky Way may arise from the 
superposition of different wave-like effects \citep{blandhawthorn2021snail}; 
we refer to \citet{Gardner:2021ntg} for 
further discussion.

We view the observed wave-like patterns in stellar number counts with position 
and 
spirals in position and velocity space as likely having shared origins, even if 
the particular observational data sets employed in the two sorts of studies are quite different --- and we will use  
``wave-like patterns'' to refer to position-space structures henceforth. 
The corrugations, or radial waves, 
noted by \citet{xu2015rings}, 
become ring- or shell-like
structures still farther from the Sun and Galactic mid-plane. These
stellar overdensities could have an accretion origin, 
via a tidally disrupted satellite 
galaxy~\citep{searle1978composition}, 
but they could also have 
come from stars ejected from the disk~\citep{xu2015rings}, with 
further observational studies supporting 
that latter, novel  
interpretation~\citep{price2015reinterpretation,li2017exploring,sheffield2018disk}. This, in turn, 
has set the stage for broader studies 
tying the global response of the 
Galactic disk to the local 
disturbances we have noted. 
For example, 
stellar kinematics measurements from {\it Gaia}
have been used 
to map asymmetric features of
the Galactic disk~\citep{katz2018gaia,drimmel2022gaia}. 
We also note evidence that 
its asymmetries may be tied to its large-scale
spiral structure~\citep{levine2006spiral,eilers2020strength,poggio2021galactic}, 
as well as various theoretical developments: namely, 
that of a systematic 
theoretical framework for the study of
dynamical phase spirals~\citep{banik2022comprehensive} 
and of numerical simulations 
of the collision of a Sagittarius-like dwarf
galaxy with the Milky Way. The latter body of work 
acts 
to discern 
its 
local and global kinematics signatures~\citep{hunt2021resolving}, 
the
vertical response of 
the disk~\citep{poggio2021measuring}, 
and the existence of various snail-like 
features~\citep{Gandhi2022snailacrossscales} that
can emerge in such a 
context. 

Recent studies have relied heavily on the use of stellar velocity information 
to tease out important dynamical effects, and 
our neglect of such information might therefore seem a limitation. Rather, we emphasize that our studies
are complementary, in that we gain in sensitivity 
not only through the sheer size and quality, in terms of completeness
and precision parallax information, of the data set 
we 
have chosen, but also on our reliance on 
symmetry-breaking effects to boost the 
visibility of subtle non-steady-state effects. 
Here we develop a 2PCF analysis of the stars near the Sun to give sharpened
insights into the structure and nature of the perturbations on the 
stars within 3 kpc of the Sun. As observations with {\it Gaia} continue, we expect that 
studying the 2PCF for stars with particular velocity selections away from the mid-plane region, but 
loosely in the manner of \citet{kamdar2020spatialclustering}, will yield 
even more 
discriminating insights.

We conclude our introduction with a brief sketch of the sections to follow. 
In Sec.~\ref{sec:Theory} we develop the theory of the 2PCF for galactic dynamics, 
computing it under steady-state conditions with and without spherical symmetry, to 
establish that visible effects necessarily come from non-steady-state effects. 
In Sec.~\ref{sec:ModelModel} we describe the extensive control studies we
have made to ensure that the 
selections we make of the {\it Gaia} DR2 data can be robustly interpreted in terms of 
physical, rather than systematic, effects. 
Because we are appreciative that studies of the nature of dark matter spurs
interest in structure at the shortest distance scales~\citep{buckley2018gravitational,Gardner:2021ntg}, we carefully 
delineate our systematic limitations in resolving small-scale structures in 
Sec.~\ref{sec:resolve}. In Sec.~\ref{sec:Data} we discuss our data selection, 
based on \citet{HGY20}, and note the control we have over observational systematic
errors. Finally, in Sec.~\ref{sec:Analysis} we report all of our 2PCF results, consider
their possible origin in Sec.~\ref{sec:origin}, 
and offer a final summary and outlook in Sec.~\ref{sec:Conclusion}.

\section{Theory} \label{sec:Theory}

\subsection{The 2PCF in steady-state}
An isolated galaxy 
is described by a  
distribution function (DF)
in its stars in six-dimensional phase space: 
$f(\mathbf{v},\mathbf{x}, t)$, with different DFs for different
stellar populations possible. 
A DF can be self-consistently determined by the 
solution of the collisionless Boltzmann, or Vlasov, 
and Poisson equations, where we emphasize that the Vlasov equation itself emerges 
only if correlations between the stars are neglected~\citep{binney2008GD}.
In steady state, such a galaxy with a stellar disk is expected to be 
axially symmetric with respect to rotations about an axis, 
through its center of mass, perpendicular to the plane of the disk, 
and thus is also reflection symmetric about the galactic mid-plane~\citep{an2017reflection,schutz2017disklimit}.
We have determined that in our own Galaxy, however, that even if axial
symmetry is very nearly conserved,  
reflection symmetry can be 
markedly broken~\citep{GHY20,HGY20}, implying that the Galaxy 
is not isolated and/or not in steady state. We interpret the small 
axial symmetry breaking we have found in our carefully selected 
sample of {\it Gaia} DR2 stars as arising, in part, 
from the net torque exerted on our 
sample within the Galaxy by the massive LMC/SMC system, yet
the differences in axial symmetry breaking we find, comparing North with South, 
are much larger still. Thus we think our results are particularly indicative
of the presence of non-steady-state effects~\citep{GHY20,HGY20}. This and the appearance
of striking wave-like features in stellar number counts
North and South of the Galactic place~\citep{widrow2012galactoseismology,yanny2013stellar,bennett2018vertical} 
suggest that the stars are likely correlated as well, possibly on many 
different length scales~\citep{kamdar2021preliminaryclusteringindisk}.
To explore this concretely, we revisit the derivation of 
the Vlasov equation itself:
we return to the Bogoliubov, Born, Green, Kirkwood, and Yvon (BBGKY) hierarchy 
which comes from the analysis of 
Liouville's equation in the 
presence of pairwise forces, 
relating the time-evolution of the $s$-particle distribution function 
$f_s$, 
to the $(s+1)$-particle distribution function $f_{s+1}$~\citep{Gardner:2021ntg}. 
Consequently, the $s$-particle distribution function 
is not simply proportional to $(f)^s$; rather, 
we introduce~\citep{thorne2017MCP}
\begin{equation}
f_2(\mathbf{v}_{1},\, \mathbf{x}_{1},\, \mathbf{v}_{2},\, \mathbf{x}_{2},\, t) 
= f_1(\mathbf{v}_{1},\, \mathbf{x}_{1},\, t)f_1(\mathbf{v}_{2},\, \mathbf{x}_{2},\, t) (1 + \xi_{12})\,,
\label{ftwo}
\end{equation}
where $\xi_{12}$ is the 2PCF, 
with the same arguments
as that of the two-particle distribution function. In general, the 
2PCF 
can either 
enhance or suppress the joint probability of finding a particle in a 
phase-space volume $d^3\mathbf{x}_1 d^3\mathbf{v}_1$ and another in 
$d^3\mathbf{x}_2 d^3\mathbf{v}_2$: the probability of finding one is now 
no longer independent of the probability of finding the other. 
The BBGKY hierarchy links $f_2$ to $f_3$, where 
\begin{equation}
f_3(\mathbf{v}_{1},\, \mathbf{x}_{1},\, \mathbf{v}_{2},\, \mathbf{x}_{2},\, 
\mathbf{v}_{3},\, \mathbf{x}_{3},\,
t) 
= f_1(\mathbf{v}_{1},\, \mathbf{x}_{1},\, t)f_1(\mathbf{v}_{2},\, \mathbf{x}_{2},\, t) 
f_1(\mathbf{v}_{3},\, \mathbf{x}_{3},\, t)
(1 + \xi_{12} + \xi_{13} + \xi_{23} + \xi_{123})\,,
\label{fthree}
\end{equation}
with $\xi_{ij} \equiv
\xi_{ij}(\mathbf{x}_i,\mathbf{v}_i,\mathbf{x}_j,\mathbf{v}_j,t)$.
If we neglect the possibility of the 
{\it three-particle} correlation function $\xi_{123}$, 
we can 
extract 
a single differential equation for the two-particle correlation functions.
Exact expressions for $\xi_{12}$ exist in simplified contexts, as in, e.g., \citet{Kirkwood1950JChPh..18.1040K}. 
Of particular interest to us is the case 
of electrons in an unmagnetized, thermalized plasma, as long-range forces
are present.
There 
the protons can be treated as 
a stationary background, so that the distribution functions are in 
electron degrees of freedom only, with correlations arising from electron-electron interactions --- and external forces can be neglected~\citep{thorne2017MCP}. 
We adapt this analysis to the particular case of stars in steady-state within a few kpc of the Sun, albeit the two problems differ in fundamental aspects, as we shall see. 
Since the gravitational interaction between any two stars 
does not depend on their velocities, we might 
also expect the 2PCF in this case to be velocity independent and depend only
on the inter-particle separation;  
$x_{ij}\equiv | \mathbf{x}_i - \mathbf{x}_j |$; 
we explore this possibility through explicit calculation.
To our knowledge, this is the first time such an extension has been 
explored, and we lay out the connection to statistical physics carefully. 
Returning to the BBGKY hierarchy, we note the lowest two equations for
a $N$-particle system are of
form~\citep{kardar2007SM} 
\begin{eqnarray}
&&\frac{\partial f_1}{\partial t} - \mathbf{\nabla}_{\mathbf{x}_1} \Phi_{\rm ext}\cdot \mathbf{\nabla}_{\mathbf{v}_1} f_1 + 
\mathbf{v}_1 \cdot \mathbf{\nabla}_{\mathbf{x}_1} f_1
 = \int d^3 \mathbf{x}_2 d^3 \mathbf{v}_2 \,
\nabla_{\mathbf{x}_1} \Phi_{12} \cdot \mathbf{\nabla}_{\mathbf{v}_1} f_2 
\,,
\label{bbgkyf1}
\end{eqnarray}
and
\begin{eqnarray}
&&\frac{\partial f_2}{\partial t} 
- \mathbf{\nabla}_{\mathbf{x}_1} \Phi_{\rm ext} \cdot \mathbf{\nabla}_{\mathbf{v}_1} f_2
- \mathbf{\nabla}_{\mathbf{x}_2} \Phi_{\rm ext} \cdot \mathbf{\nabla}_{\mathbf{v}_2} f_2 
+ \mathbf{v}_1 \cdot \mathbf{\nabla}_{\mathbf{x}_1} f_2 +
  \mathbf{v}_2 \cdot \mathbf{\nabla}_{\mathbf{x}_2} f_2 \nonumber \\
&& - \Big[\mathbf{\nabla}_{\mathbf{v}_1}f_2 \cdot \nabla_{\mathbf{x}_1}\Phi_{12}
 + \mathbf{\nabla}_{\mathbf{v}_2}f_2 \cdot \nabla_{\mathbf{x}_2}\Phi_{12}
 \Big]
= \int d^3 \mathbf{x_3} d^3 \mathbf{v}_3 
\Big[
\mathbf{\nabla}_{\mathbf{v}_1} f_3 \cdot \nabla_{\mathbf{x}_1}\Phi_{13}
+ 
\mathbf{\nabla}_{\mathbf{v}_2} f_3 \cdot \nabla_{\mathbf{x}_2}\Phi_{23}
\Big]\,,
\label{bbgkyf2}
\end{eqnarray}
where $\Phi_{\rm ext} (\mathbf{x}_i)$ is the external gravitational potential 
and $\Phi_{ij} (x_{ij})$ is the two-body gravitational potential, where we employ dimensions of 
energy per mass throughout~\citep{binney2008GD}. In this context it is conventional to normalize
$f_1(\mathbf{x}, \mathbf{v},t)$ so that $\int d^3 \mathbf{x} d^3 \mathbf{v} f_1 = {\cal M}$, the total mass
of the system, rather than unity. We thus adjust the RHS of Eqs.(\ref{bbgkyf1},\ref{bbgkyf2}) accordingly in what follows.
The collisionless Boltzmann equation follows by neglecting correlations, replacing 
$f_2$ with $(f_1)^2$ in Eq.~(\ref{bbgkyf1}), yielding 
\begin{eqnarray}
\frac{\partial f_1}{\partial t} - \mathbf{\nabla}_{\mathbf{x}_1} \Phi_0 \cdot \mathbf{\nabla}_{\mathbf{v}_1} f_1 + 
\mathbf{v}_1 \cdot \mathbf{\nabla}_{\mathbf{x}_1} f_1 =0 
\label{vlasovf1}
\end{eqnarray}
with 
\begin{eqnarray}
\Phi_0 = \Phi_{\rm ext} + 
\frac{1}{\cal{M}}\int d^3 \mathbf{x}_2 d^3 \mathbf{v}_2\, f_1(\mathbf{x}_2,\mathbf{v}_2, t) \Phi_{12}
=\Phi_{\rm ext} + \frac{1}{\cal{M}}\int d^3 \mathbf{x}_2 \,\rho(\mathbf{x}_2,t) \Phi_{12}
\,. 
\label{phiefff1}
\end{eqnarray}
The mass density $\rho$ and the effective potential $\Phi_0$, which are both 
regarded as smooth distributions~\citep{binney2008GD} because we assume 
$N\gg 1$, can be determined 
self-consistently by solving Eq.~(\ref{vlasovf1}) and the Poisson equation 
\begin{equation}
\nabla^2 \Phi_0 = 4\pi G \rho    \,, 
\label{poisson}
\end{equation}
with the equilibrium solution satisfying
$\partial f_1/\partial t =0$. 
Turning to the analysis of the $f_2$ equation, Eq.~(\ref{bbgkyf2}), and assuming
$f_1$ satisfies Eqs.~(\ref{vlasovf1}-\ref{poisson}), with $f_1$ and $\xi_{ij}$ in steady state, we find
\begin{eqnarray}
&&\!\!\!\!\!f_1(\mathbf{x}_1, \mathbf{v}_1) f_1(\mathbf{x}_2, \mathbf{v}_2)\left[ 
- \mathbf{\nabla}_{\mathbf{x}_1} \Phi_{0} \cdot \mathbf{\nabla}_{\mathbf{v}_1} 
- \mathbf{\nabla}_{\mathbf{x}_2} \Phi_{0} \cdot \mathbf{\nabla}_{\mathbf{v}_2}  
+ \mathbf{v}_1 \cdot \mathbf{\nabla}_{\mathbf{x}_1} +
  \mathbf{v}_2 \cdot \mathbf{\nabla}_{\mathbf{x}_2} 
\right] \xi_{12} \nonumber \\
&&\!\!\!\!\! - \left[\nabla_{\mathbf{x}_1}\Phi_{12} \cdot \mathbf{\nabla}_{\mathbf{v}_1}
 + \nabla_{\mathbf{x}_2}\Phi_{12} \cdot \mathbf{\nabla}_{\mathbf{v}_2} \right] 
 f_1(\mathbf{x}_1, \mathbf{v}_1 ) f_1(\mathbf{x}_2, \mathbf{v}_2 )(1  + \xi_{12})
 \nonumber \\
&&\!\!\!\!\!=\frac{1}{\cal{M}}\int d^3 \mathbf{x}_3 d^3 \mathbf{v}_3 
\Big[ \nabla_{\mathbf{x}_1} \Phi_{13} \cdot 
\mathbf{\nabla}_{\mathbf{v}_1} 
+ \nabla_{\mathbf{x}_2} \Phi_{23} \cdot
\mathbf{\nabla}_{\mathbf{v}_2} 
\Big]f_1(\mathbf{x}_1, \mathbf{v}_1 ) f_1(\mathbf{x}_2, \mathbf{v}_2 ) f_1(\mathbf{x}_3, \mathbf{v}_3 ) (\xi_{12} + \xi_{13} + \xi_{23}) \label{bbgkyx12} 
\,
\end{eqnarray}
--- noting we have set $\xi_{123}=0$. 
In what follows we analyze two simple cases. Although neither one offers a 
realistic description of the Milky Way galaxy, we find our analytic analysis 
of value in that it 
shows clearly that the parametric behavior of the 2PFC in steady state is 
grossly different from what we observe in the data. 
In the first we suppose that 
$f_1$ is that of an infinite, isotropic, spherically symmetric system, 
for which
\begin{equation}
  f_1 (v) = \frac{\rho_0}{(2\pi \sigma^2)^{3/2}} \exp\left(- \frac{v^2}{2\sigma^2}\right)  \,, 
  \label{sphf1}
\end{equation}
where $\rho_0$ is a constant. 
This 
is not a self-consistent model because only
Eq.~(\ref{vlasovf1}) is satisfied; here we commit the  ``Jeans swindle'' and set $\Phi_0=0$~\citep{binney2008GD}.
In the second case we suppose that $f$ depends on the vertical coordinate
only, giving a slab geometry. For concreteness we employ the self-consistent thin-disk
model of \citet{Spitzer1942ApJ....95..329S} in this latter case.
In both cases we assume that 
$\xi_{ij}$ is independent of velocity, and we justify this assertion
{\it a posteriori}, from the form of the solutions we find. 
Since $\Phi_{ij} = -GM/|\mathbf{x}_i - \mathbf{x}_j|$, where $M$ is the
mass of an $\epsilon$-sphere centered on $\mathbf{x}_j$, 
$M=\int_{|\mathbf{x}'| \le |\mathbf{x}_j| + \varepsilon} d^3\mathbf{x}' \rho_0(\mathbf{x}')$, nominally the mass of a star. 
Thus 
$\nabla_{\mathbf{x}_i}\Phi_{ij}= - \nabla_{\mathbf{x}_j}\Phi_{ij}$ 
for any $i\ne j$, 
and Eq.~(\ref{bbgkyx12}) becomes
\begin{eqnarray}
&& (\mathbf{v}_1 \cdot \mathbf{\nabla}_{\mathbf{x}_1})\xi_{12}
+ (\mathbf{v}_2 \cdot \mathbf{\nabla}_{\mathbf{x}_2})\xi_{12}
+ \frac{1}{\sigma^2} (1 + \xi_{12}) (\mathbf{v}_1 - \mathbf{v}_2 ) \cdot \mathbf{\nabla}_{\mathbf{x}_1}\Phi_{12}
 \nonumber \\
&& = - \frac{\rho_0}{{\cal M}\sigma^2} 
\int d^3 \mathbf{x_3} \,
(\xi_{13} + \xi_{23})
\Big[  
(\mathbf{v}_1 \cdot 
 \mathbf{\nabla}_{\mathbf{x}_1} ) \Phi_{13}
+ 
(\mathbf{v}_2 \cdot \mathbf{\nabla}_{\mathbf{x}_2} )\Phi_{23} 
\Big]
\,.
\label{bbgkyx12_etc}
\end{eqnarray}
The velocity dependence is explicit, and the equation must 
hold for any choice of $\mathbf{v}_1$ and $\mathbf{v}_2$. 
Setting $\mathbf{v}_2 =0$ for simplicity, and assuming 
$\xi_{12} \ll 1$, we determine 
\begin{equation}
\mathbf{\nabla}_{\mathbf{x}_1} \Big(\sigma^2 \xi_{12} 
+ \Phi_{12} \Big) = - \frac{\rho_0}{\cal{M}} \int d^3 \mathbf{x}_3 \,
(\xi_{13} + \xi_{23}) \mathbf{\nabla}_{\mathbf{x}_1} \Phi_{13} \,.
\end{equation}
Since $\mathbf{\nabla}_{\mathbf{x}_1} \Phi_{13}$ changes sign if 
$\mathbf{x}_1 \stackrel{>}{{}_<} \mathbf{x}_3$, we see if $\xi_{13}$ depends only 
on $x_{13}$ its contribution to the integral vanishes. Assuming this and 
with $\mathbf{\nabla}_{\mathbf{x}_1}^2 \Phi_{12} 
= 4 \pi G M \delta^{(3)}(\mathbf{x}_{12})$, we finally have 
\begin{equation}
\mathbf{\nabla}_{\mathbf{x}_1}^2 \xi_{12} 
+ \frac{4\pi \rho_0 G M}{{\cal M}\sigma^2} \xi_{12} = - \frac{4\pi G M}{\sigma^2} \delta^{(3)}(\mathbf{x}_{12}) \,.
\end{equation}
This reveals that $\xi_{12}$ is the response of the system to the interparticle
interaction sourced at $\mathbf{x}_2$. 
Introducing $\lambda_{\rm G} \equiv \sqrt{{\cal M}\sigma^2/4\pi G M \rho_0}$, 
we first solve this equation for $x_{12}\ne 0$ to yield 
\begin{equation}
    \xi_{12} (x_{12}) = 
    A \frac{\cos(x_{12}/\lambda_{\rm G})}{x_{12}} + B\frac{\sin(x_{12}/\lambda_{\rm G})}{x_{12}}\,,
    \label{x12gfirst}
\end{equation}
where $A$ and $B$ are arbitrary constants. Through consideration of 
the 
divergence theorem in the $\rho_0\to 0$ limit we find 
\begin{equation}
    \xi_{12} (x_{12}) = \left(\frac{G M}{\sigma^2} \right)
    \frac{\cos(x_{12}/\lambda_{\rm G})}{x_{12}} \,,
    \label{x12grav}
\end{equation}
The attractive nature of the gravitational interaction
dictates the form of our solution. In contrast, 
the unmagnetized, thermal electron gas has a
2PCF of form
$\xi_{12} \propto \exp(- x_{12}/\lambda_{\rm D})/x_{12}$, where $\lambda_{\rm D}$ is the Debye length~\citep{thorne2017MCP}.
It is useful to contrast our result with the encounter operator~\citep{binney2008GD}, namely,
\begin{equation}
    \Gamma[f(\mathbf{x}_1,\mathbf{v}_1,t)] 
    \equiv \frac{\cal{M}}{M} \int d^3 \mathbf{x_2} d^3 \mathbf{v}_2 \mathbf{\nabla}_{\mathbf{x}_1} \Phi_{12} \cdot \mathbf{\nabla}_{\mathbf{v}_1} \xi_{12} \,,
\end{equation}
which drives the change of the (one-body) DF, 
\begin{equation}
\frac{df}{dt} = \Gamma[f] \,.
\end{equation}
Employing our result in Eq.~(\ref{x12grav}) we see that the 
right-hand side does indeed evaluate to zero, so that our 2PCF 
is compatible with a steady-state limit. 
It is worthwhile to study the explicit length scales
associated with our solution. We see that 
Eq.~(\ref{x12grav}) can attain a value of ${\cal O}(1)$ 
or larger if and only if 
$x_{12} <  G M/\sigma^2$. Supposing 
$\sigma$ to be given by the vertical velocity dispersion of the disk, 25 ${\rm km \ s^{-1}}$ \citep{minchev2014new},
$M\approx M_\odot = 1.988\times 10^{30}\, \rm kg$, 
and 
$G=6.674\times 
10^{-11}\,\rm m^3 \ kg^{-1} \ s^{-2}$ \citep{PDG2020}, 
we determine that $ G M/\sigma^2 \approx 
2.1 \times 10^{11}\, \rm m = 
6.9\times 10^{-6}\,\rm pc$. 
Thus for the length scales of interest to us, 
the BBGKY hierarchy should give a reasonable estimate of
the size of the correlation effects in steady state, and $\xi_{12}$ is also extremely 
small. We note that $\lambda_G$ can be written as $\lambda_G= \sqrt{\sigma^2 V_{\rm eff}/4\pi GM}$,
where $V_{\rm eff}$ is the effective volume per star. Referring to the extremely
complete {\it Gaia} DR2 sample of \citet{HGY20}, within roughly $3\,\rm kpc$ of the Sun, we note that 
11.7 million stars occupies a volume of at least $16\,{\rm kpc}^3$ 
to yield $\lambda_G\gtrsim \sqrt{f_{\rm col/mag}}(4.0 \,{\rm kpc})$, where $f_{\rm col/mag}$ is
the fraction of the total number of stars that fall within the chosen 
color and magnitude cuts. 
Querying the {\it Gaia} DR2 database we 
determine $f_{\rm col/mag}\approx 0.3$, so that $\lambda_G \gtrsim 2.2\,{\rm kpc}$.
With $x_{12} = 0.20\,\rm kpc$, e.g., we find that Eq.~(\ref{x12grav}) 
evaluates to $\xi_{12} = 3.4\times 10^{-8}$. Thus 
it would appear that any nonzero values of 
$\xi_{12}$ that we would be able to observe
cannot come from steady-state effects. 
The spherical symmetry of $f_1$ dictates that 
$\xi_{12}$ in Eq.~(\ref{x12grav}) can only depend on 
the scalar $x_{12}$. To consider how our results might change with the
symmetries of the problem, we now turn to our second example, that of 
a slab geometry, with $f_1$ depending on 
the vertical energy $E_z$ only, i.e., 
\begin{equation}
  f(z,v) = \frac{\rho_0}{\sqrt{2\pi \sigma^2}} \exp(-E_z/\sigma^2)   \,,
  \label{fspitzer}
\end{equation}
where $E_z = v^2/2 + \Phi_0(z)$, 
$\Phi_0(z)=2 \sigma^2 \ln ({\rm cosh}(z/2z_d))$, and $\rho_0=\sigma^2/8\pi G z_d^2$~\citep{Spitzer1942ApJ....95..329S}.
In this case, $\Phi_{ij}$ represents the potential at $z_i$ due to a 
infinitely thin and uniform sheet of surface mass density $\Sigma$ at $z_j$, so that 
$\Phi_{ij} = 2\pi G\Sigma |z_{i} - z_{j}|$.
Returning to Eq.~(\ref{bbgkyx12}), we note that {\it if $\xi_{12}$ is independent
of velocity} that the terms in $\Phi_0$ do not contribute, and we 
find that $\xi_{12}$ satisfies
\begin{eqnarray}
&& (v_1 \partial_{z_1} + v_2 \partial_{z_2} 
)\xi_{12}
+ \frac{1}{\sigma^2} (1 + \xi_{12}) (v_1 - v_2 ) \partial_{z_1}
\Phi_{12}
 \nonumber \\
&& = - \frac{1}{{\Sigma}\sigma^2} 
\int dz_3 \,
\rho(z_3)
(\xi_{12} + \xi_{13} + \xi_{23})
\Big[  
v_1 \partial_{z_1} \Phi_{13} 
+ 
v_2 \partial_{z_2} \Phi_{23} 
\Big]
\,,
\label{bbgkyx12_zdep}
\end{eqnarray}
where $\rho(z)=\rho_0 {\rm sech}^2 (z/2z_d)$ and the
surface mass density $\Sigma=\int_{-\infty}^{\infty} dz\,\rho(z)=4\rho_0 z_d$.
Supposing that 
$\xi_{ij}$ depends in some manner on $|z_i - z_j|$ as well, we see that 
\begin{equation}
    \int dz_3 \,
\rho(z_3) \xi_{i3} \partial_{z_3} \Phi_{i3} \approx 0 \,;  \int dz_3 \,
\rho(z_3) \partial_{z_3} \Phi_{i3} \approx 0
\end{equation}
for $i=1,2$ 
if $z_1, z_2$ are in the vicinity of the Galactic mid-plane. 
Thus 
 \begin{eqnarray}
&& (v_1 \partial_{z_1} + v_2 \partial_{z_2} 
)\xi_{12}
+ \frac{1}{\sigma^2} 
(1 + \xi_{12}) 
(v_1 - v_2 ) \partial_{z_1}
\Phi_{12}
 \nonumber \\
&& = - \frac{1}{{\Sigma}\sigma^2} 
\int dz_3 \,
\rho(z_3)
\Big[  v_1 \xi_{23} \partial_{z_1} \Phi_{13} 
+ 
v_2 \xi_{13} \partial_{z_2} \Phi_{23} 
\Big]
\,.
\label{bbgkyx12_zdep_etc}
\end{eqnarray}
Proceeding as in 
the previous case, we find 
\begin{equation}
\partial_{z_1} \Big(\xi_{12} 
+ \frac{1}{\sigma^2} \Phi_{12} \Big) = - \frac{1}{{\Sigma}\sigma^2} \int d {z}_3 \,
\rho(z_3) \xi_{23} {\partial_{{z}_1}} \Phi_{13} \,,
\end{equation}
and since
\begin{equation}
   \partial_{z_1}^2 \Phi_{12} = 4\pi G \Sigma_{z_2} \delta(z_1 -z_2) \,,
\end{equation}
where we introduce $\Sigma_{z_2}$ as 
the surface mass density of a thin and uniform sheet
at $z_2$, with $\Sigma_{z_2}=\Delta\rho(z_2)$ and a parameter $\Delta$ 
of ${\cal O}(z_d)$, we find 
\begin{equation}
\partial_{z_1}^{2} \xi_{12} 
+
\frac{4\pi G\rho(z_1) \Sigma_{z_1}}{{\Sigma}\sigma^2} \xi_{12} = 
- \frac{4\pi G \Sigma_{z_1}}{\sigma^2} \delta(z_{1} - z_{2}) 
\label{slabxi12_1}
\end{equation}
for $z_1,z_2$ close to the Galactic mid-plane, noting 
$\Sigma_{z_1} \ll \Sigma $. 
Since explicit $z_1$ dependence 
appears in the second term of Eq.~(\ref{slabxi12_1}), which might be expected
because the matter distribution is not isotropic, 
we see that 
$\xi_{12}$ cannot depend on $|z_1-z_2|$ alone. 
Thus we introduce $z_{12}=z_1 - z_2$ and $Z=(z_1 + z_2)/2$ to find 
\begin{equation}
(\partial_{z_{12}}^{2} + \partial_{z_{12}} \partial _{Z} + \frac{1}{4} \partial_{Z}^2 )\xi_{12} 
+
\frac{4\pi G\rho(z_{12}/2 + Z) \Sigma_{z_{12}/2 + Z}}{{\Sigma}\sigma^2} \xi_{12} = 
- \frac{4\pi G \Sigma_{z_{12}/2 + Z}}{\sigma^2} \delta(z_{12}) \,, 
\label{slabxi12_2}
\end{equation}
and finally 
\begin{equation}
\partial_{z_{12}}^{2} 
\xi_{12} 
+
\frac{4\pi G\Delta \rho_0^2  }{{\Sigma}\sigma^2} \xi_{12} = 
- \frac{4\pi G \rho_0 \Delta}{\sigma^2} \delta(z_{12}) \,, 
\label{slabxi12_3}
\end{equation}
near the Galactic mid-plane. 
With 
$\lambda_{\tilde {\rm G}} \equiv \sqrt{{\Sigma}\sigma^2/4\pi G \Delta \rho^2_0}$, 
we solve this equation for 
$z_{12}\ne 0$ to yield 
\begin{equation}
    \xi_{12} (z_{12}) = 
    A \cos(z_{12}/\lambda_{\tilde{\rm G}})
    + B\sin(z_{12}/\lambda_{\tilde{\rm G}})
    \,,
    \label{slabx12_4}
\end{equation}
where $A$ and $B$ are arbitrary constants. 
Considering the $z_d \to \infty$ limit with $\Delta$ fixed, we
see that only the term with $A$ survives and thus we estimate
\begin{equation}
    \xi_{12} (z_{12}) 
    \approx  
    A 
    \left(1 - \frac{z_{12}^2}{2\lambda^2_{\tilde {\rm G}}} \right)
    \,
    \label{slabx12_soln}
\end{equation}
in the mid-plane region 
--- and we note that $\xi_{12}$ is symmetric
under $z_1 \leftrightarrow z_2$ exchange.
In this case we are unable to fix 
the strength of the homogeneous solution through consideration of the 
source term and a suitable Gaussian surface, 
but we observe that variations in $\xi_{12}$ are determined by 
$\lambda_{\tilde {\rm G}}$, which we evaluate to be
$\lambda_{\tilde {\rm G}} 
\gtrsim \sqrt{\sigma^2/\pi G \rho_0} = \sqrt{\sigma^2 4z_d / \pi G \Sigma}$ 
for 
$\Delta \lesssim z_d$. Using the surface mass density in stars at $|z|=1.1\,\rm kpc$ from 
\citet{bovy2013direct}, namely $38 \pm 4 \,\rm M_\odot pc^{-2}$, and a scale 
height of $280 \,\rm pc$~\citep{bovy2015galpy} we determine 
$\lambda_{\tilde {\rm G}} \gtrsim 1.1 \,\rm kpc$, which is within a factor of
2 of our estimate in the purely isotropic case. We thus conclude that the 
spatial variations associated with $\xi_{12}$ in the steady-state case
are 
roughly comparable to 
the physical dimensions
of our stellar sample, 
though in the regions in which Eq.~\ref{slabx12_soln} would be valid,
the spatial correlations we would be able to 
observe cannot come from steady-state effects. 
In this regard we emphasize that our analytic solutions
for $\xi_{12}$, namely, Eqs.~(\ref{x12grav},\ref{slabx12_soln}), 
do not break the underlying symmetries present in each case. 
This also supports our notion that symmetry breaking 
speaks to the appearance of non-steady-state effects. 
In the next section we determine how the 2PCF can be determined from 
observations.

\subsection{Evaluation of the 2PCF: connecting theory to observations}
In this paper, we wish to access the 2PCF of the Milky Way 
in an entirely data-driven way. We do so by adapting the 2PCF analysis 
in galactic number counts familiar from the study of 
cosmic large-scale structure~\citep{Peebles1993ppc..book.....P}
and thus begin by reviewing that setting, referring to 
\citet{Peebles1993ppc..book.....P} 
for all details. 
The joint probability of finding two galaxies, which is assumed to be stationary, at separation 
$r$ centered within volume elements $dV_1$ and $dV_2$, respectively, is 
$dP_2 = n^2 (1 + \xi(r/r_0)) dV_1 dV_2$, where the probability to find one
galaxy is $dP_1=ndV$. In this context $\xi(r/r_0)$, the two-point correlation function
depends on $r_0$, a characteristic clustering length, 
which is determined from observations.
If the universe is homogeneous, we 
can convert this quantity to an angular correlation function that can be directly 
determined 
from the data by including a selection function $S_i$ which determines the likelihood
that a galaxy $i$ at some distance is bright enough to be detected. With this, 
the joint probability becomes 
\begin{equation}
    dP_2 = n^2 d\Omega_1 d\Omega_2 \int r_1^2 dr_1 r_2^2 dr_2 \left(1 + \xi(r_{12}/r_0)\right)
    S_1 S_2 \,, 
\end{equation}
and thus we have 
\begin{equation}
    dP_2 = N^2 d\Omega_1 d\Omega_2 (1 + w(\theta)) \,,
\end{equation}
with 
\begin{equation}
w(\theta) = \frac{\int r_1^2 dr_1 r_2^2 dr_2 \xi(r_{12}/r_0) S_1 S_2}{(\int r^2 dr S)^2}\,,
\end{equation}
where $r_{12} = (r_1^2 + r_2^2 - 2 r_1 r_2 \cos \theta)^{1/2}$ and $N$ is the mean 
number of galaxies per steradian. 
In order to assess $w(\theta)$ from the observational data, 
the Landy-Szalay (LS) estimator \citep{landy1993bias} is employed, though other choices
are possible~\citep{wall2012practical}. In this method the data $D$ with 
$d$ points is compared to a reference
model $R$ with $r$ points, which is comprised of randomly distributed galaxies, 
and three separate histograms are constructed: 
$RR$, $DD$, and $DR$.  
Each histogram counts the number of pairs of stars at separations of 
$\theta$ to $\theta+d\theta$, and 
$DD$ counts these pairs using the data, $RR$ counts them within the reference model, and 
$DR$ counts the number of cross-correlation pairs, to yield 
\begin{equation}
    w_{\rm LS}(\theta) = \frac{RR(\theta) - 2DR(\theta) + DD(\theta)}{RR(\theta)} \,,
\end{equation}
where the histograms must be suitably normalized. 

Segueing to our Milky Way studies, 
we first note that our selected {\it Gaia} DR2 set is 
exceptionally 
complete in our selected color and magnitude windows~\citep{HGY20}, 
and we have not applied 
a selection 
function as a result. 
We note that complete velocity information is only available 
for stars which are brighter than those in our sample, and the 
gravitational interaction does not depend on velocity, 
so that we consider the 
density-density correlation function and do not restrict
 the relative velocities of the stars in any way. 
 The matter distribution in the Milky Way is not 
 spherically symmetric, so that we expect the 
 joint probability of finding two stars at separation $\mathbf{x}_{12} \equiv \mathbf{x}_1 - \mathbf{x}_2$
 to depend on the projection onto the Cartesian vectors 
 $\hat{e}_x$, $\hat{e}_y$,and $\hat{e}_z$ as well as on 
 $\mathbf{x}_1$, or $\mathbf{x}_2$, itself.
 For the stars of our sample
 within a fixed region of the sky, we expect the sample-averaged 2PCF to be determined by 
\begin{equation}
    \langle \xi_{12}(|\hat{e}_Q\cdot \mathbf{x}_{12}| ) \rangle 
    \equiv 
    \left\langle 
    \rho (\mathbf{x}_i) \rho (\mathbf{x}_i + \mathbf{x}_{12})
    \right\rangle_{{\rm fixed}\, |\hat{e}_Q\cdot \mathbf{x}_{12}|}\,,
    \label{x12defn}
\end{equation}
for a fixed choice of ${Q} \in x, y, z$, where the
average is determined by summing over the coordinate $\mathbf{x}_i$ of
each of the 
stars in our 
selected sample. We employ standard practice, so that 
$\mathbf{x}$ points along the anti-Center line towards
the Galactic center, and $\mathbf{y}$ points in the 
direction of $\phi< 180^\circ$. 
In concrete terms we use 
the LS 
estimator~\citep{landy1993bias}, with histograms formed by the 
histograms that count the number of pairs of stars separated by 
some distance $q=|\hat{e}_Q\cdot \mathbf{x}_{12}|$ 
to $q+dq$ with $dq=d|\hat{e}_Q\cdot \mathbf{x}_{12}|$ 
(and henceforth $z_{12} > 0$, e.g.).
In usual practice, $DD$ would count pairs 
within our {\it Gaia} data set, $RR$ would count pairs within a 
reference theoretical model, sampled statistically, and 
$DR$ would count the cross-correlation pairs between the two data sets to yield 
\begin{equation}
    \langle \xi_{\rm LS}(q_i) \rangle = \frac{RR(q_i) - 2DR(q_i) + DD(q_i)}{RR(q_i)} \,, 
    \label{eq:x12_LS_MW}
\end{equation}
where we have averaged over the coordinates not fixed by $q_i$. 
As noted previously, 
the $RR$, $DD$, and $DR$ histograms must be
normalized so that they have unit areas \citep{wall2012practical}.  
This normalization accounts for the 
potentially different numbers of stars in real and mock samples, 
as well as 
the fact that there are more cross-correlation pairs than pairs within one data set. 

Here we differ from usual practice, as we wish to 
exploit the near symmetry of our chosen data set to tease out the 2PCF 
arising from non-steady-state effects. 
That is, we choose ``$D$'' and ``$R$'' to be  
two distinct selections of the {\it Gaia} data sample of \citet{HGY20}, that  
are related by either axial or North-South reflection symmetry, 
rather than 
comparing the observational data with 
a model-dependent mock catalogue. 
For example, 
one can examine the difference in structure between the Northern hemisphere of the Galaxy and (a reflection of) the Southern hemisphere, so long as the geometries of the two regions are identical. This is enabled through the approximate reflection symmetry of the Galaxy, 
to yield a new and sensitive probe of the symmetries the Milky Way.

\section{Control Studies:  Interpreting the 2PCF} \label{sec:ModelModel}

Regardless of whether the LS estimator is used in a traditional manner (i.e., comparing model vs. data) or to probe symmetry-breaking effects as in this work (i.e., comparing data vs. data), it is critical to understand how the LS estimator reveals structure in order to (i) avoid being tricked by the effects of the various geometric cuts, the $z_{\odot}$ offset, etc. 
and (ii) to understand how effects, such as North-South symmetry breaking, 
observed in the data vis-a-vis asymmetries in the 
one-body density across the Galactic plane, 
are manifested in a two-point correlation study.  
In this section we demonstrate that comparing two samples drawn from suitably 
chosen models, with 
explicit symmetries in place, 
suffices as probes of both of these issues, 
because an explicitly symmetric model can serve as either a model to 
compare with data {\it or} as a simulation of reflected data.
To this end, we have simulated a number of different scenarios and 
detail them in this section.  

\begin{figure}[h!]
  \begin{center}
    \subfloat[]{\includegraphics[scale=0.55]{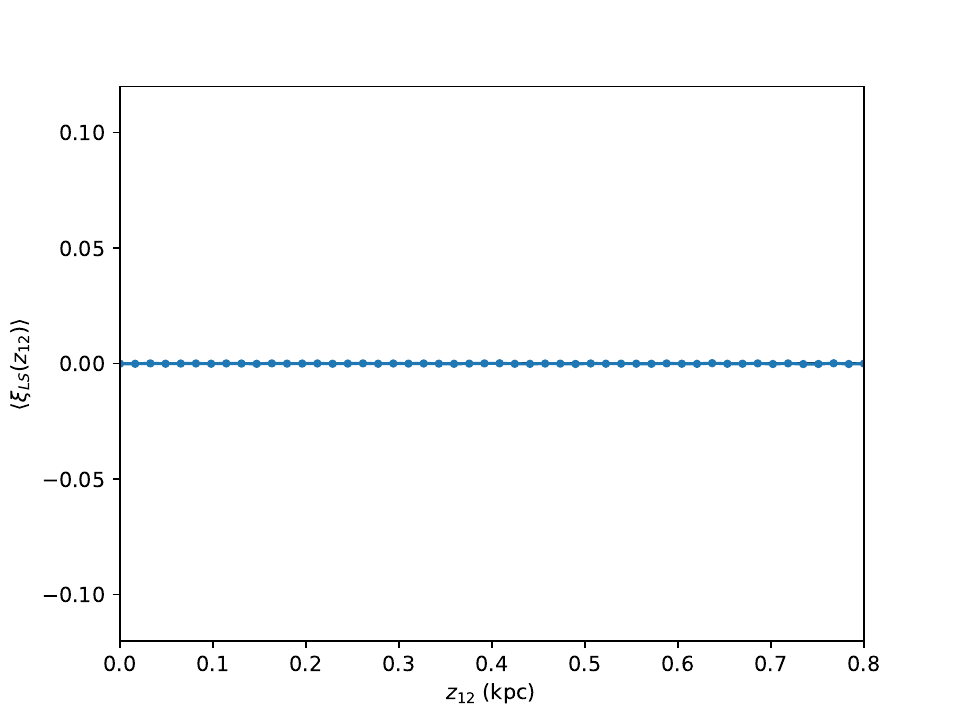}}
    \subfloat[]{\includegraphics[scale=0.55]{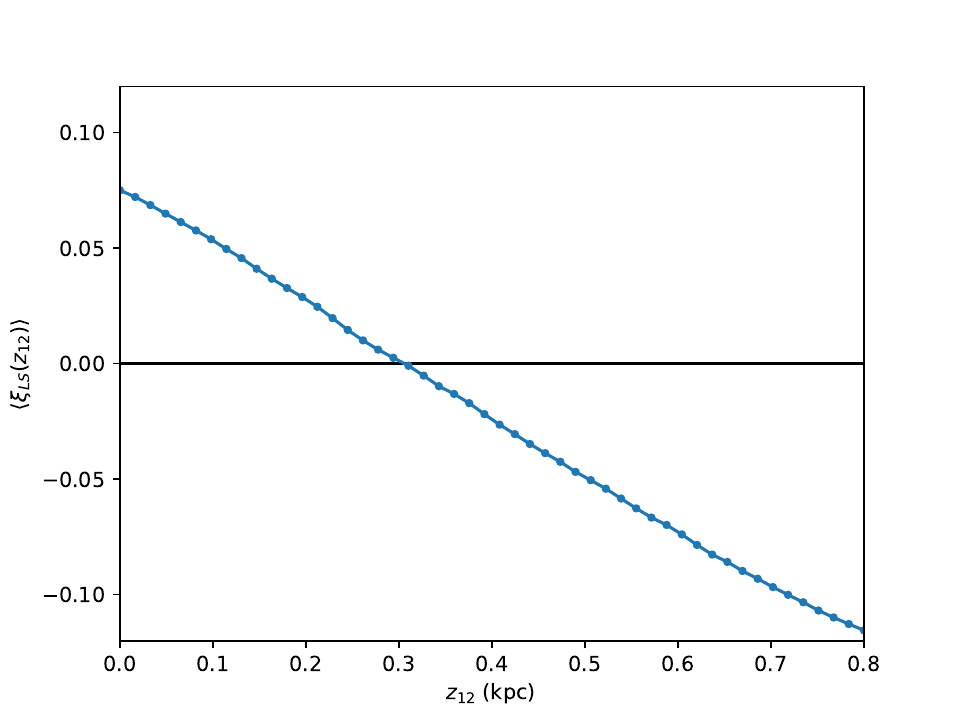}}

    \subfloat[]{\includegraphics[scale=0.55]{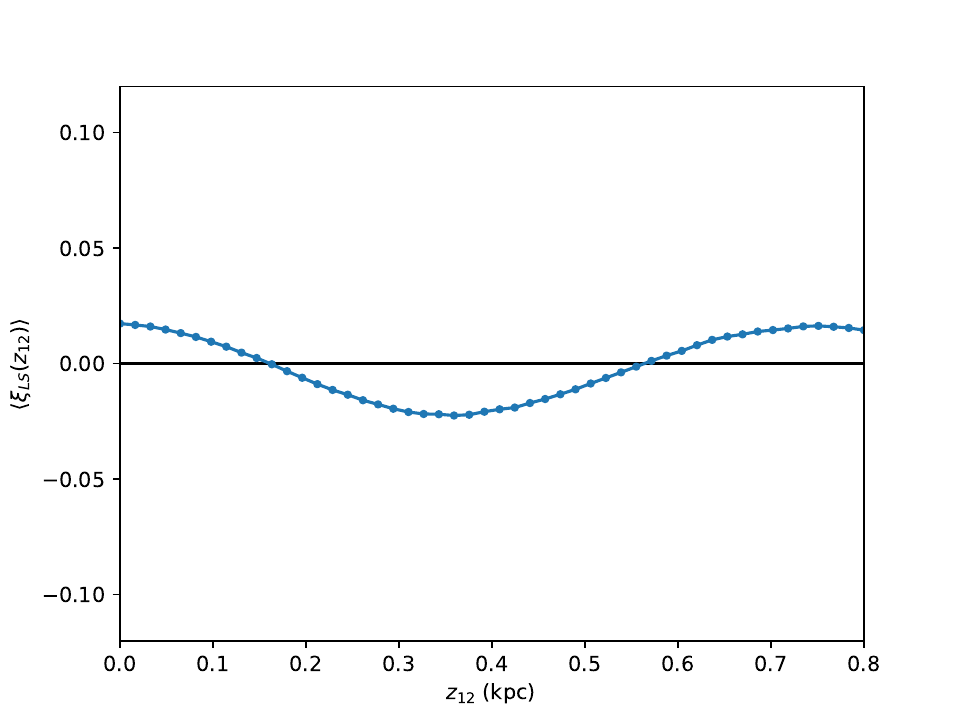}}
    \subfloat[]{\includegraphics[scale=0.55]{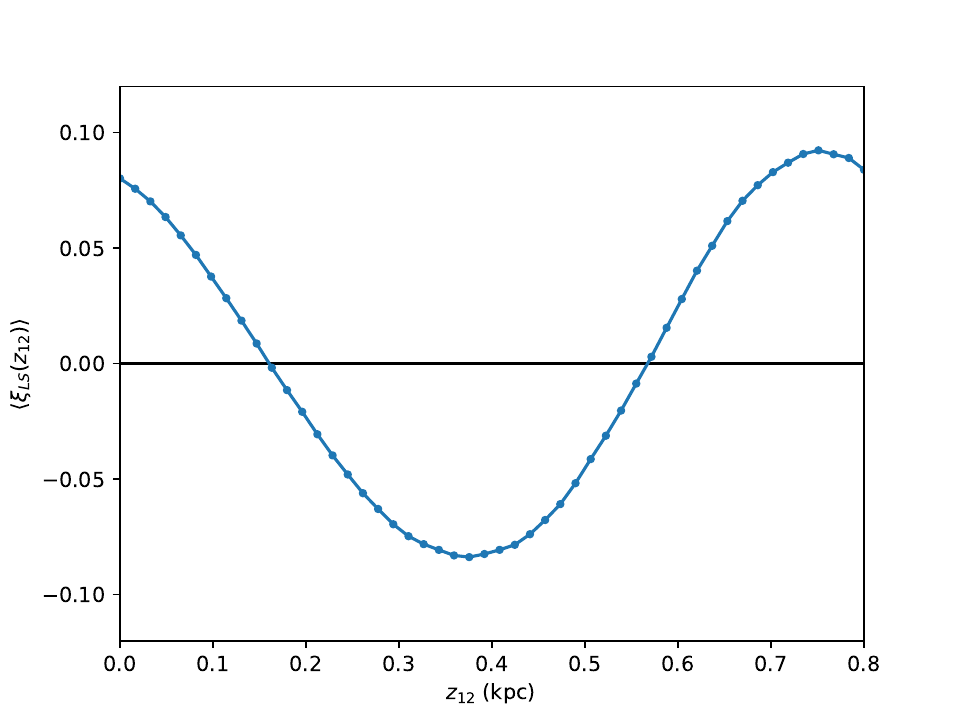}}
    \caption{
    (a) The N vs. S z2PCF comparing two models drawn from identical distribution functions. 
    (b) The N vs. S z2PCF comparing two models drawn from distribution functions with different scale heights.  In this case, one model has a scale height of $z_{\rm s} = 280$ pc, while the other has a scale height of $z_{\rm s} = 420$ pc.  
    (c) The z2PCF comparing a model with vertical density waves against a smooth model. 
    (d) The N vs. S z2PCF comparing a model with vertical density waves in the North against the same model with vertical density waves in the South.  The wave is modelled as in Eq.~(\ref{Eq:wave})
    and thus the anti-symmetric nature of the wave results in more significant correlations in the North vs. South analysis than that comparing the density wave model against a smooth model.
    All panels implement the following cuts: $7.6 < R < 8.2$ kpc, $176 < \phi < 184^{\circ}$, $0.2 < |z| < 2.0$ kpc. 
    }
  \label{fig:modelModel1}
  \end{center}
\end{figure}

In Fig.~\ref{fig:modelModel1}(a), the LS estimator is shown for $z$-separation distances, and it compares two data sets drawn from an identical distribution function, of form ${\rm sech}^2 (z/2z_s)$.  This control test illustrates the case where no structure exists, because
the two samples are drawn from identical distributions, 
and indeed the LS estimator is consistent with zero.  A result with structure is shown in Fig.~\ref{fig:modelModel1}(b), wherein two data sets are compared with markedly different scale heights.  The first data set ($D$) has a scale height of $z_{\rm s} = 280$ pc \citep{bovy2015galpy}, while the other ($R$) has an intentionally inflated scale height of $z_{\rm s} = 420$ pc, and thus an excess of structure is found at small scales, and a dearth of structure is found at larger scales, resulting in the slanted estimator shown.

If instead of a smooth, hyperbolic-secant-squared distribution function, we use a DF with structure embedded in it, the LS estimator will then pick out the characteristic scales of that particular structure.  For example, by introducing a toy-model of a vertical wave into one data set ($D$) as in Eq.~(\ref{Eq:wave}), 
\begin{equation}
    n(R, z) = e^{\sfrac{-R}{R_{\rm s}}} {\rm sech}^2\left( \frac{z}{2z_{\rm s}} \right)\left( 1 + 0.2 \cdot {\rm sin}(8z) \right)
    \label{Eq:wave}
\end{equation}
and comparing this against smoother data drawn from a hyperbolic-secant-squared distribution function ($R$), 
both with the same $z_s=280\, \rm pc$, the LS estimator will indicate an excess of structure corresponding to the maxima of the density waves, and a dearth of structure near the trough of the density waves, as in Fig.~\ref{fig:modelModel1}(c).  An interesting feature of an anti-symmetric structure like that of the vertical waves found in \citet{widrow2012galactoseismology, yanny2013stellar, bennett2018vertical} is that any North-South comparison will result in increased significance in a structure search.  In other words, when data in the North ($DD$) is compared against a reflection of the data in the South ($RR$), the anti-symmetric nature of the vertical waves results in the peaks in the North lining up with the troughs of the South, and  thus the LS estimator strongly highlights this structural difference, as depicted in Fig.~\ref{fig:modelModel1}(d).

\begin{figure}[h!]
  \begin{center}
    \subfloat[]{\includegraphics[scale=0.55]{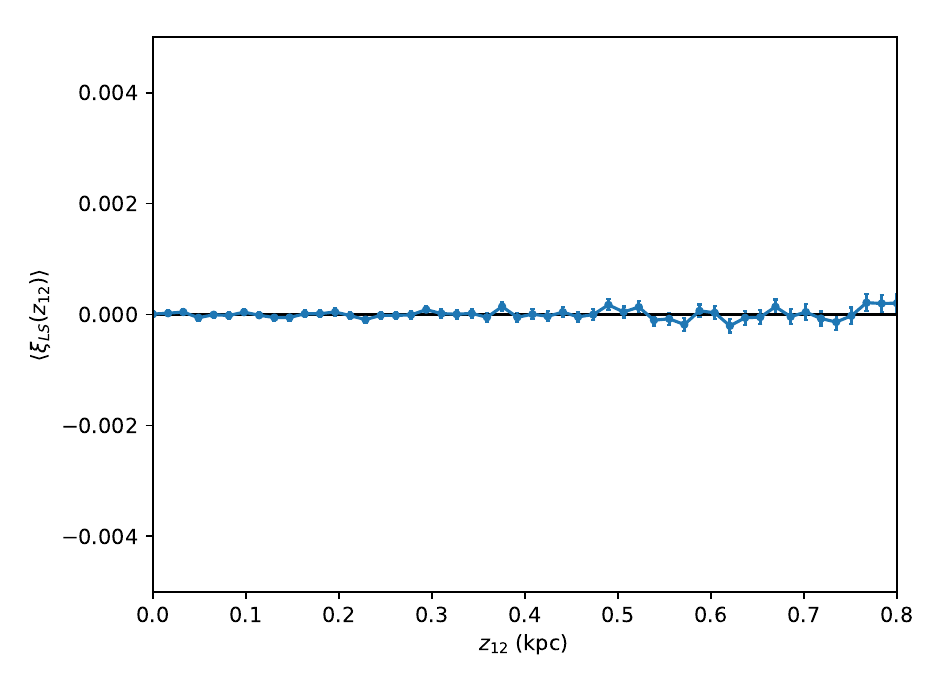}}
    \subfloat[]{\includegraphics[scale=0.55]{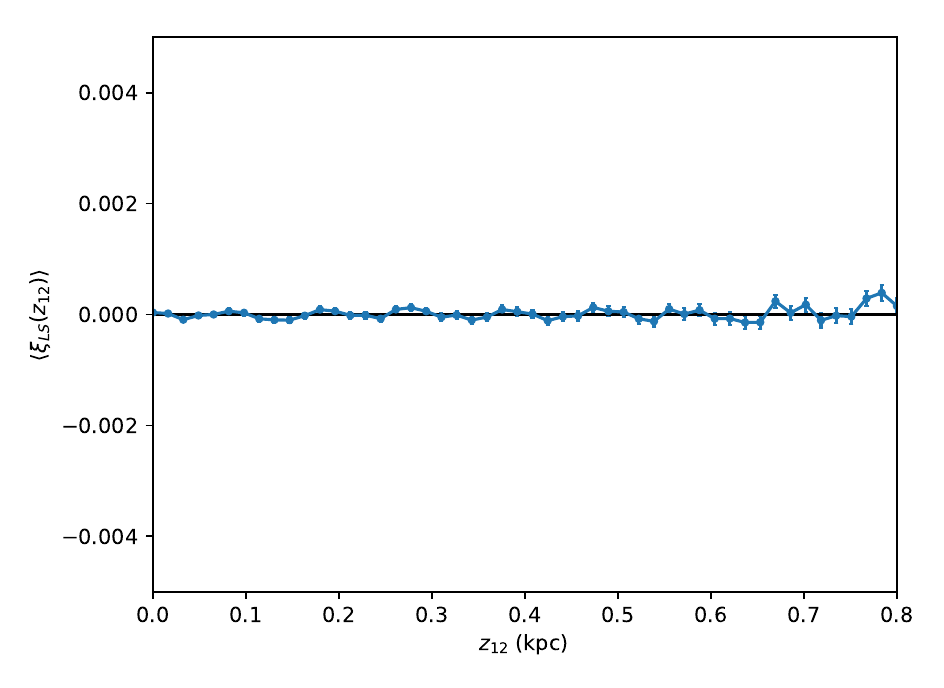}}
    
    \subfloat[]{\includegraphics[scale=0.55]{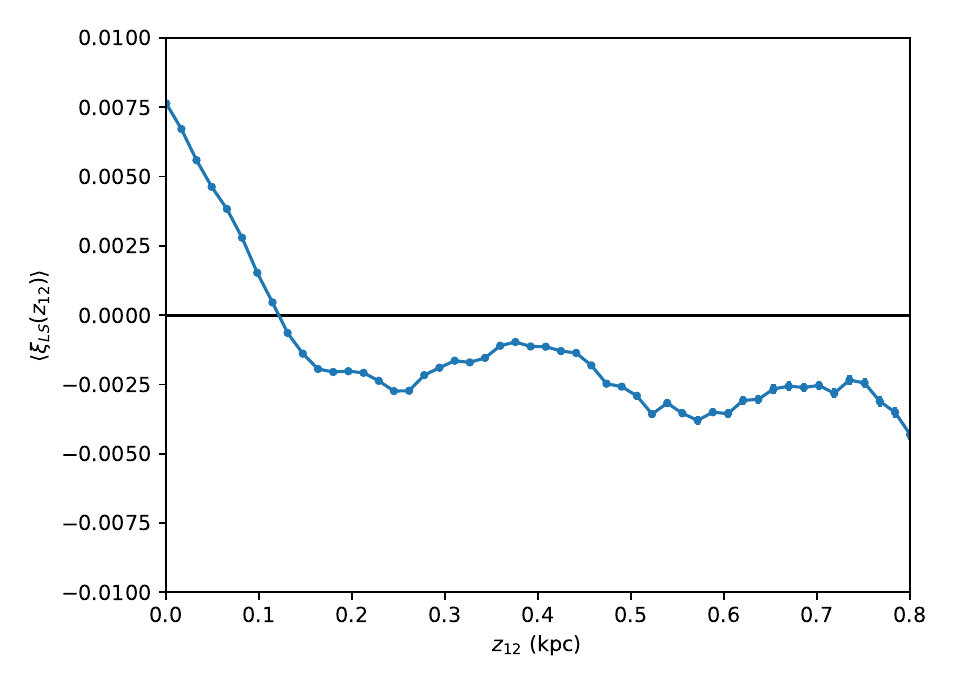}}
    \subfloat[]{\includegraphics[scale=0.55]{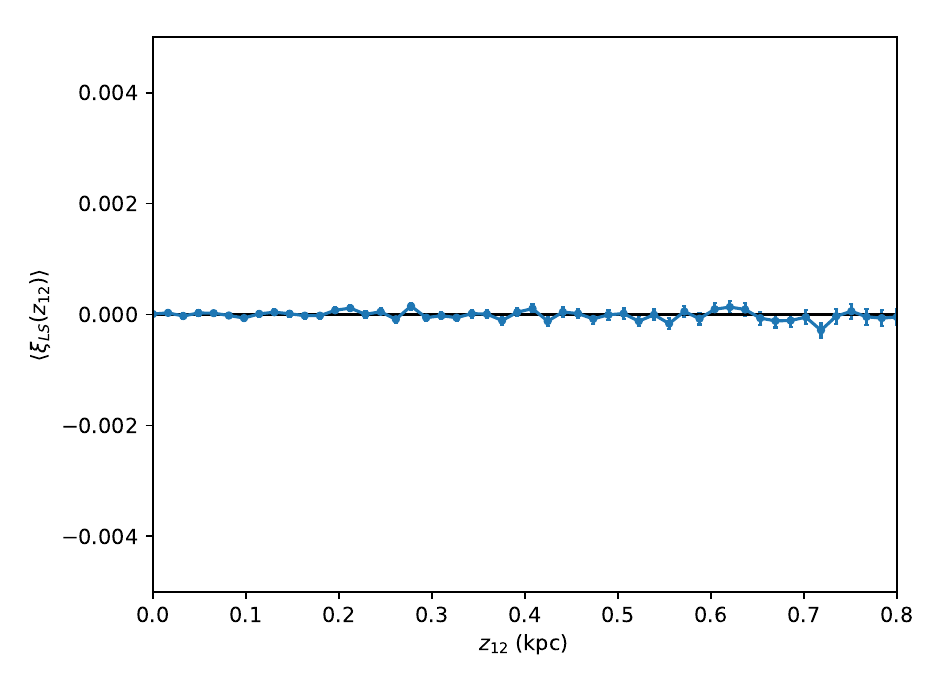}}
    \caption{
    (a) The N vs. S z separation 2PCF comparing two models drawn from identical distribution functions in the region $7.6 < R < 8.2$ kpc, $176^{\circ} < \phi < 184^{\circ}$, $0.2 < |z| < 2.0$ kpc, including $|b| > 30^{\circ}$ cuts and the set of LMC/SMC cuts mentioned in the text.
    (b) The same scenario as panel a, but with a 20\% mismatch in the number of stars between the two models.
    (c) The N vs. S z2PCF comparing two models drawn from a distribution function which accounts for $z_{\odot}$, resulting in a geometry mismatch, North and South, because 
    the heliocentric cuts 
    no longer emanate from mid-plane.
    (d) The N vs. S z2PCF comparing two models drawn from a distribution function which accounts for $z_{\odot}$, but with Galactocentric geometry chosen in such a way that the $l$ and $b$ cuts are avoided.  In this particular example, $7.9 < R < 8.3$ kpc, $179^{\circ} < \phi < 181^{\circ}$, and $0.3 < |z| < 2.0$ kpc, and thus the $|b| > 30^{\circ}$ cuts and the LMC/SMC cuts do not impact the geometry of the sample, which results in no geometry mismatch from heliocentric cuts.
    }
  \label{fig:modelModel2}
  \end{center}
\end{figure}

While  
helpful for illustrative purposes, the toy models studied in  Fig.~\ref{fig:modelModel1} are missing a key consideration.  In selecting a reliable data set as free from observational artifacts as possible, we have implemented various cuts on {\it heliocentric} longitude and latitude.  However, because the sun is not truly situated on the Galactic mid-plane, any analysis in galactocentric coordinates will necessarily run into problems caused by a geometry mismatch. To better illustrate this concept, let us consider Fig.~\ref{fig:modelModel2}.  In panel (a), we repeat the control test of Fig.~\ref{fig:modelModel1}(a), but now include cuts on latitude ($|b| > 30^{\circ}$) as well as the LMC and SMC cuts of \citet{GHY20, HGY20}.  In this case, a $z_{\odot}$ shift has not been applied, and thus these toy models implicitly assume $z_{\odot} = 0$ pc.  It is clear from panel (a) that $l$ and $b$ cuts alone do not bias the LS estimator if $z_{\odot} = 0$ pc.  Indeed, even if the samples have an egregiously large mismatch in the number of stars, as in Fig.~\ref{fig:modelModel2}(b), the LS estimator still takes into account both the post-cut geometry as well as normalization considerations, correctly resulting in no indication of structure. 
However, if $z_{\odot} \neq 0$, where we convert heliocentric to galactocentric
coordinates via $z\to z + z_\odot$, 
a galactocentric analysis will incur substantial geometric effects in the LS estimator, as seen in Fig.~\ref{fig:modelModel2}(c).  In this case, an offset of $z_{\odot} = 20$ pc \citep{bennett2018vertical} has been included in the models, such that heliocentric cuts on $l$ and $b$ effectively emanate from a region which is not coincident with the Galactic mid-plane.  This difference in geometries, North and South, is falsely registered in the LS estimator as structure, even though the models are identical in all other regards.  Thus, we must be extremely careful to avoid geometric differences caused by a combination  of non-zero $z_{\odot}$ and cuts on $l$ and $b$.

\begin{figure}[h!]
  \begin{center}
    \hspace*{+0.45cm} 
    \subfloat[]{\includegraphics[scale=0.43]{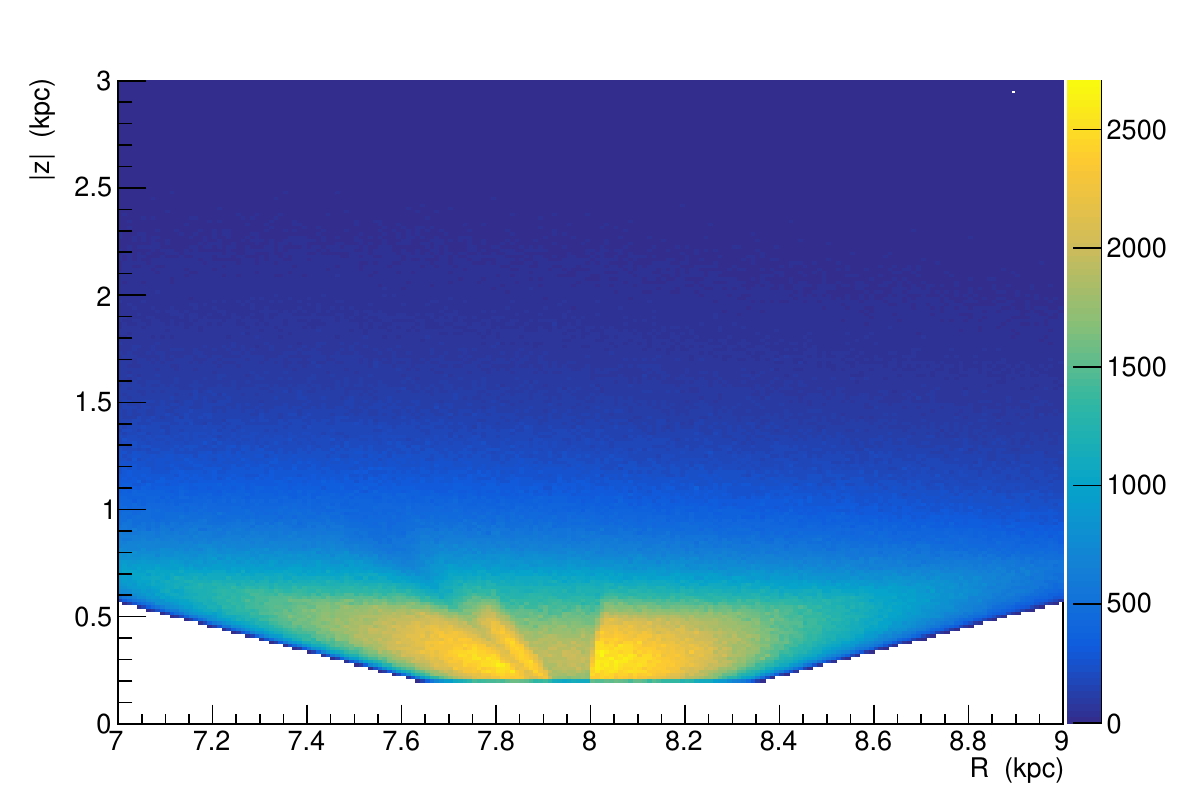}}
    \subfloat[]{\includegraphics[scale=0.43]{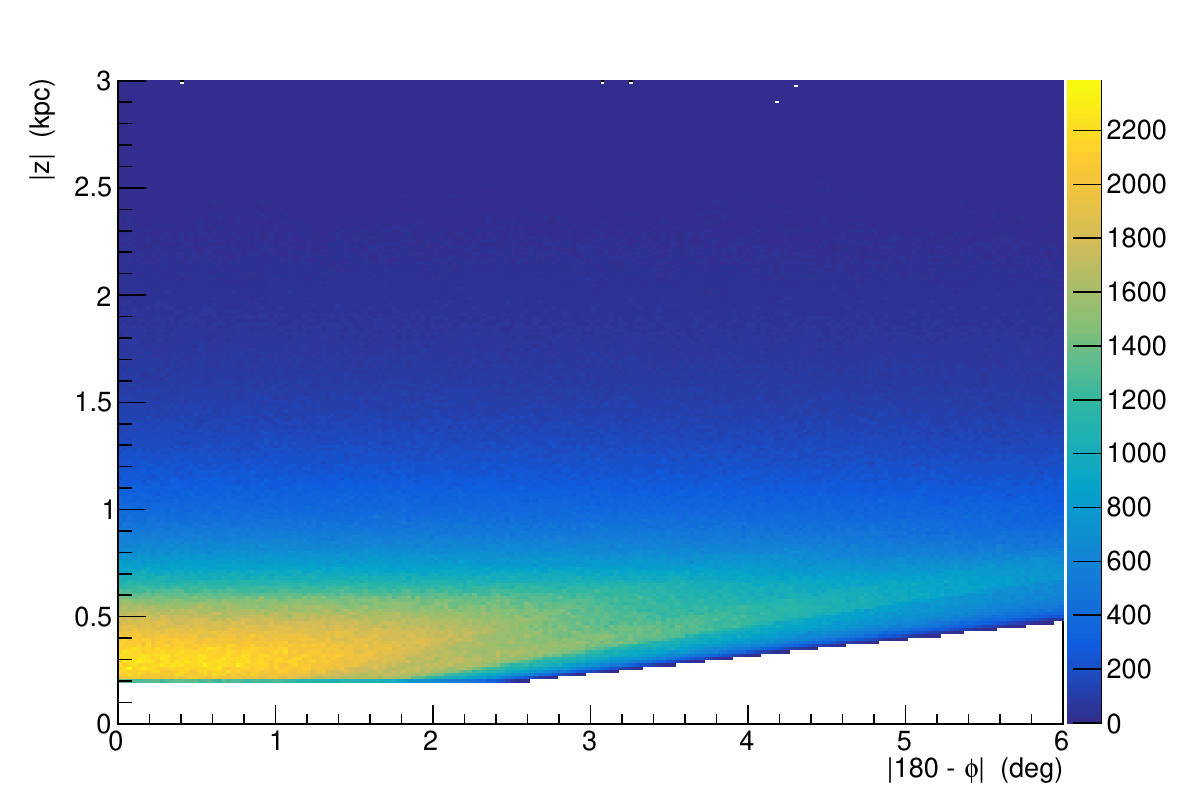}}
    
    \subfloat[]{\includegraphics[scale=0.55]{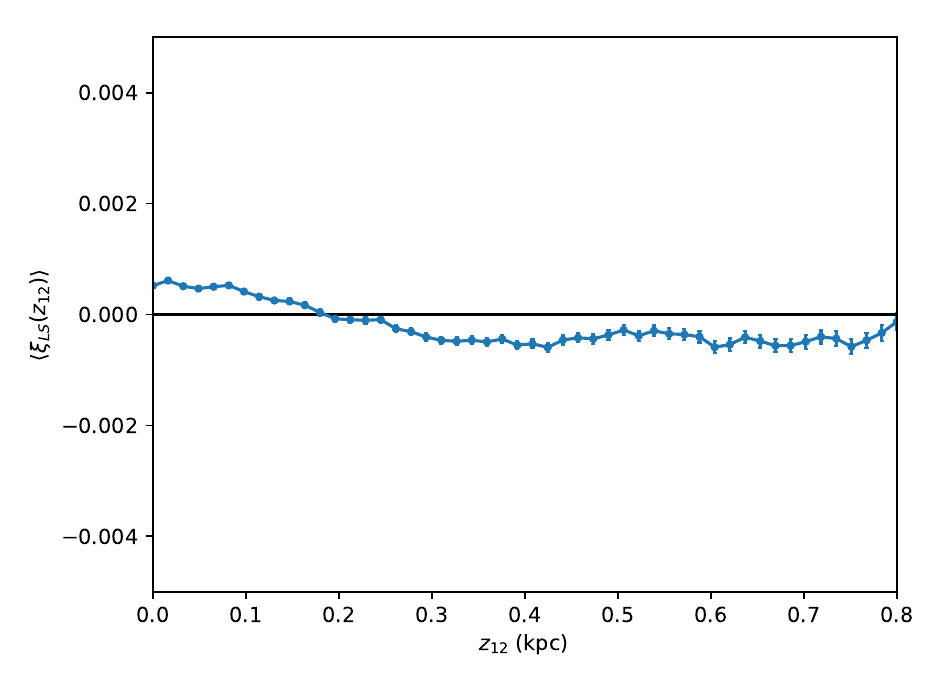}}
    \subfloat[]{\includegraphics[scale=0.55]{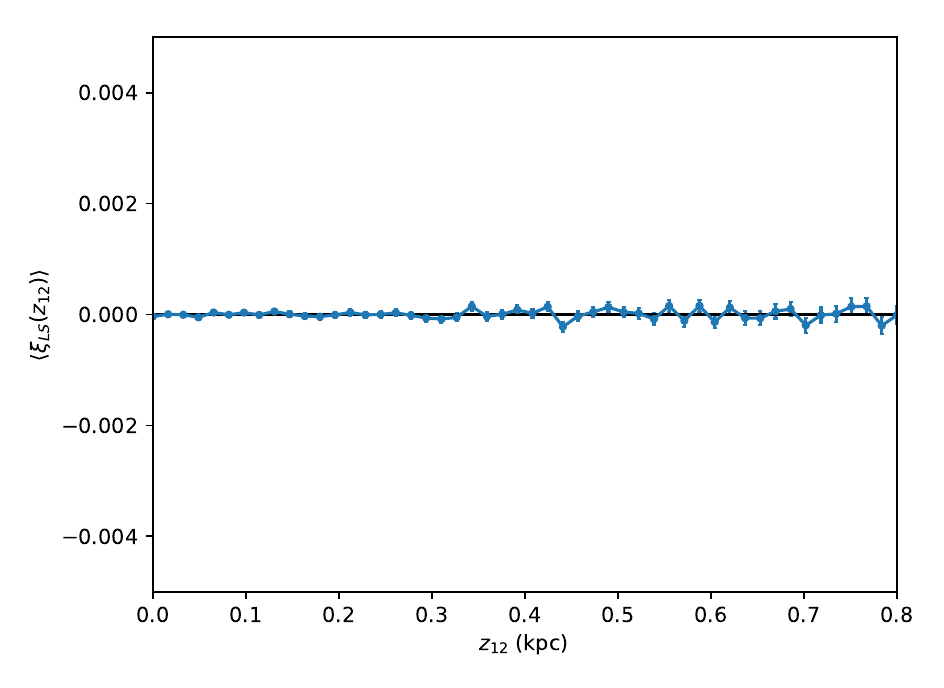}}
    \caption{
    (a) The $R-z$ projection of the data set, illustrating the regions free of $l,b$ cut effects.
    (b) The $\phi-z$ projection of the data set, illustrating the regions free of $l,b$ cut effects.
    (c) A heliocentric analysis of the 2PCF, so that it avoids 
    the geometry mismatch problem of the galactocentric analysis shown 
    in panel c of Fig.~\ref{fig:modelModel2}. Here, $7.6 < R < 8.2$ kpc, $176^{\circ} < \phi < 184^{\circ}$, $0.2 < |z| < 2.0$ kpc.  In this case the small, non-zero correlation arises due to the sampling of slightly different regions of the Galactic distribution function 
    because 
    the $z_{\odot}$ offset is {\it not} included. 
    (d) A heliocentric analysis similar to panel c, but for higher $z$:  $1.2 < |z| < 3.0$ kpc.  Here the smallest correlations 
    one can probe are 
    significantly smaller for regions well above the mid-plane,  as explained in the text.
    }
  \label{fig:modelModel3}
  \end{center}
\end{figure}

There are two potential remedies 
for 
this issue.  First, it is possible to select data such that the $l$ and $b$ cuts are avoided entirely.  In these cases, an analysis in galactocentric coordinates works without trouble, 
as the heliocentric cuts simply do not enter the geometry in question.  To illustrate this point, Fig.~\ref{fig:modelModel2}(d) shows how samples drawn from identical distributions 
that avoid the $l$ and $b$ cuts result in an LS estimator which is consistent with zero.  In this particular example, we have raised the minimum value of $z$ to which we probe, and have restricted the region of $R$  and $\phi$ as well.  While this is 
certainly a  viable workaround for the geometry mismatch issue stemming from heliocentric cuts in a Galactocentric coordinate system, the procedure limits the regions that we can explore.
To illustrate this, 
the regions of the data set impacted by the $l$ and $b$ cuts are shown via their projections in the $R-z$ and $z-\phi$ planes in Figs.~\ref{fig:modelModel3}(a) and~\ref{fig:modelModel3}(b),  respectively.  As the latter shows, increasing the minimum $z$ to which we probe is an effective way of avoiding the $l$ and $b$ cuts, at the cost of cutting out the region with the highest number of stars.  Regions at low-$z$ are still available to explore with this method in limited regions of $R$, as shown in the former panel.

An alternative 
remedy for the heliocentric cut mismatch issue is to  
conduct the analysis in heliocentric coordinates.  That is, we assume $z_{\odot} = 0$ pc.  While not strictly true, we show in Fig.~\ref{fig:modelModel3}(c) that any false correlations due to this incorrect choice of $z_{\odot}$ are small. 
In particular, the correlations we find in this case are $\xi_{\rm LS} < 0.001$, and they appear because
we are sampling slightly different regions of the Galaxy's distribution function in the North and in the South.  To wit, the 20 pc $z_{\odot}$ ``shift" is small compared to the scale height of $\sim$ 280 pc, and so the resulting effect is small, but it is nonetheless important vis-a-vis the smallest significant correlation we can probe.  

Any structure will need to exceed this background correlation of about 0.0005, as shown in Fig.~\ref{fig:modelModel3}(c), to be physically significant.  Moreover, for high $|z|$, the disk's density profile falls off approximately as an exponential decay function.  Because a shift in an exponential function is equivalent to an overall normalization factor, and because the LS estimator takes into account differences in normalization, a $z_{\odot}$ shift will not matter 
in this case.  In general, though, other distributions would 
retain some effects from a $z_{\odot}$ shift. 
We note for 
regions high above the mid-plane, neglecting $z_{\odot}$ results in negligible correlations for an otherwise identical model-model comparison, indicating the stars in our simple model do fall off approximately exponentially at high $|z|$, as shown in Fig.~\ref{fig:modelModel3}(d). 
Turning to the {\it Gaia} data, in regions not impacted by our heliocentric cuts, we have explicitly verified that the numerical choice of $z_{\odot}$ does not affect the outcomes of our study;
we regard this check as complementary to the one of  Fig.~\ref{fig:modelModel3}( c).

\section{Systematic Limitations in Resolving Small-Scale Structures}\label{sec:resolve}
Although the {\it Gaia} 
data in the solar neighborhood is remarkably complete, 
boasting an extraordinary number ($>10^9$) of stars, 
it is nevertheless the 
case that the 
distribution of stars is finite, so that 
the stars, on average, are 
separated by 
some typical length scale $\lambda_{\rm lim}$.  This density-derived limitation fundamentally limits the smallest length scale 
to which our analysis can reliably probe.

\begin{figure}[h!]
  \begin{center}
    \subfloat[]{\includegraphics[scale=0.55]{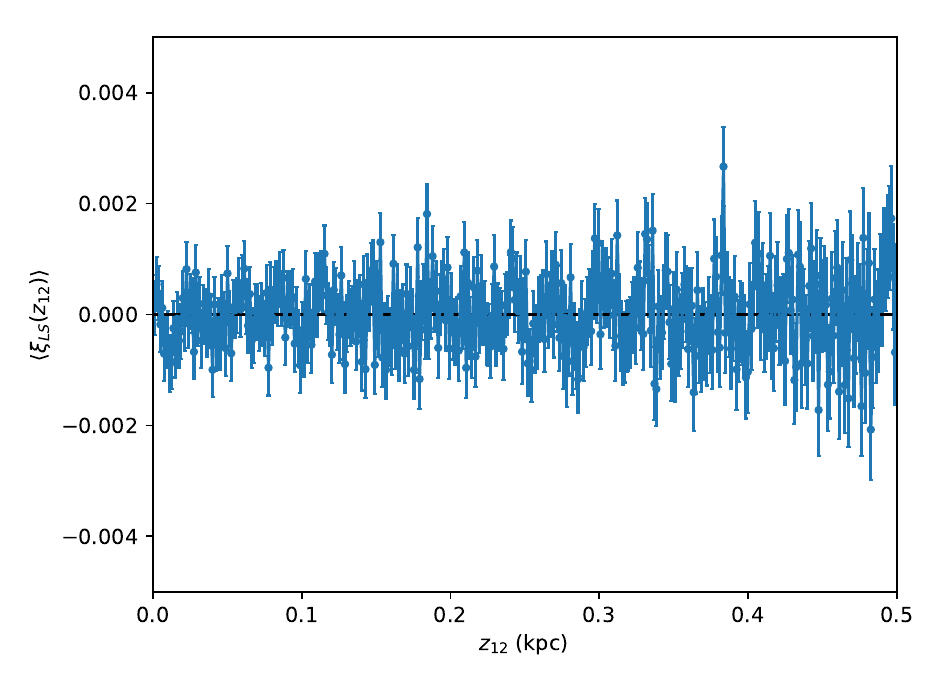}}
    \subfloat[]{\includegraphics[scale=0.55]{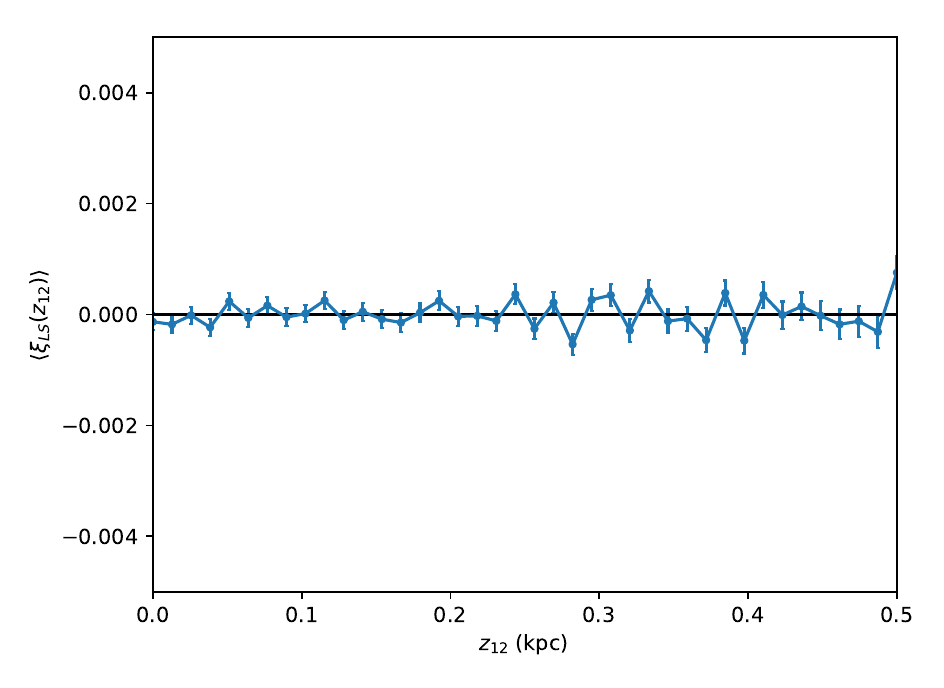}}
    
    \subfloat[]{\includegraphics[scale=0.55]{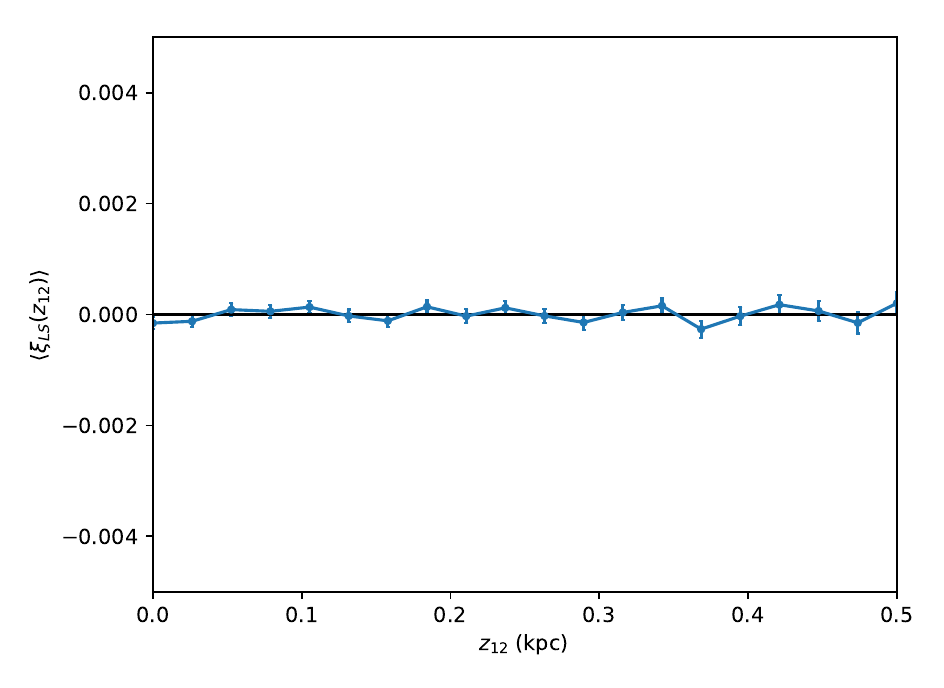}}
    \caption{ A comparison of two model Galaxies drawn from an identical distribution, as noted in text, with bin widths of (a) 2 pc, (b) 20 pc, and (c) 40 pc.  As the models were drawn from the same distribution, there should be no excess structure, yet binning too finely exposes the study to the finite density 
    limitations 
    described in the text.  The approximate limiting length scale for this example simulation is $\lambda_{\rm lim} \approx 20$ pc, and thus the apparent structure in panel (a) is not significant, while both panels (b) and (c) 
    correctly 
    indicate a lack of significant structure. 
    }
  \label{fig:smallestScale}
  \end{center}
\end{figure}

As an illustrative example of this limitation, let us examine two mock data sets drawn from the 
same distribution, of 
Gaussian 
form, centered on $z=0$ with standard deviation
$\sigma_z = 280\, \rm pc$, 
each with 100,000 stars.  These simulations were made 
in a small volume with a height of 800 pc and the z2PCF was computed to assess structural difference between the two mock galaxies.  As the galaxies were drawn from identical distributions, any structure must be due to systematic effects.  The data is binned in three different ways with 400, 40, and 20 bins as depicted in panels (a), (b), and (c) of Fig.~\ref{fig:smallestScale}, respectively.  With a vertical extent of 800 pc, these histograms of the LS estimator 
have bin widths of 2 pc, 20 pc, and 40 pc, respectively. Turning to 
Fig.~\ref{fig:smallestScale}(a), we see that binning the data too finely results in false structure, while a more coarsely binning as in panels (b) and (c) correctly indicates a lack of structure, matching expectations for two mock galaxies drawn from identical distributions.  Again, because the limiting scale length increases with decreasing density of stars, we expect some kind of 
volume-per-particle dependence.  Indeed, 
Fig.~\ref{fig:smallestScale} suggests the
limiting length scale is approximately the cube root of the volume-per-particle, or more conveniently:
\begin{equation}
    \lambda_{\rm lim} \approx \frac{L_{i}}{N^{\sfrac{1}{3}}},
    \label{eq:minlambda}
\end{equation}
where $L_{i}$ is the length of the sample in the direction in which we analyze the 
one-dimensional LS estimator and $N$ is the total number of stars. 
For example, the illustrative simulation in Fig.~\ref{fig:smallestScale} had a height of 800 pc and 100,000 stars, yielding a limiting scale of about 20 pc.  The ultra-fine binning in panel (a) is much smaller than this limiting length scale, and we observe
false structure from over-binning the data.  As panels (b) and (c) are each plotted 
with a bin size of at least $\lambda_{\rm lim}$, we see no significant structure.

The finite density effect we have considered is 
critical to interpreting the results of our LS estimator analysis. 
While bin widths will sometimes be finer than the limiting scale in what follows, only structures with scales exceeding $\lambda_{\rm lim}$ 
can be 
significant. Future data releases may possibly 
enable the study of still smaller scales, possibly even at 
the sub-Solar system level 
through consideration of planetary-scale objects. 

In addition to the finite density effects we have noted, the precision of 
our 2PCF study is limited by 
uncertainties in {\it Gaia}'s parallax measurements and thus to their distance determinations.  
Errors in distance assessments will likely ``smear" or ``blur" structure, and thus
they are unlikely to result in false structure.  
No evidence for appreciable 
direction-dependent parallax errors has been observed 
in the {\it Gaia} data, and the sample of nearby 
stars we have employed has, on average, 
relative parallax errors of 8.6\% \citep{HGY20}.  

Finally, the uncertainties in the LS Estimator are calculated from the individual uncertainties in its three component histograms.  
The $RR$, $DD$, and $DR$ histograms count pairs of stars, and thus each obey Poisson statistics.  
These individual uncertainties, assumed to be uncorrelated, are then propagated forward to quantify the uncertainty of the LS Estimator.

Altogether, when the value of the LS Estimator exceeds $1\sigma$ from zero over a scale in excess 
of $\lambda_{\rm lim}$, structure is deemed as significant. 
Additionally, horizontal error bars are not included, as distance errors are uncorrelated with location on the sky and thus not expected to create false structure given the large number of stars in the analysis (see Sec.~\ref{sec:Data}). 
With a firmer picture of the correlations 
we can probe now in place, 
we now turn to a description of our methods.

\section{Methodology and Data Selection} \label{sec:Data}
\begin{figure}[h!]
  \begin{center}
    \subfloat[]{\includegraphics[clip, trim=0.5cm 7cm 0.5cm 7cm, scale=0.4]{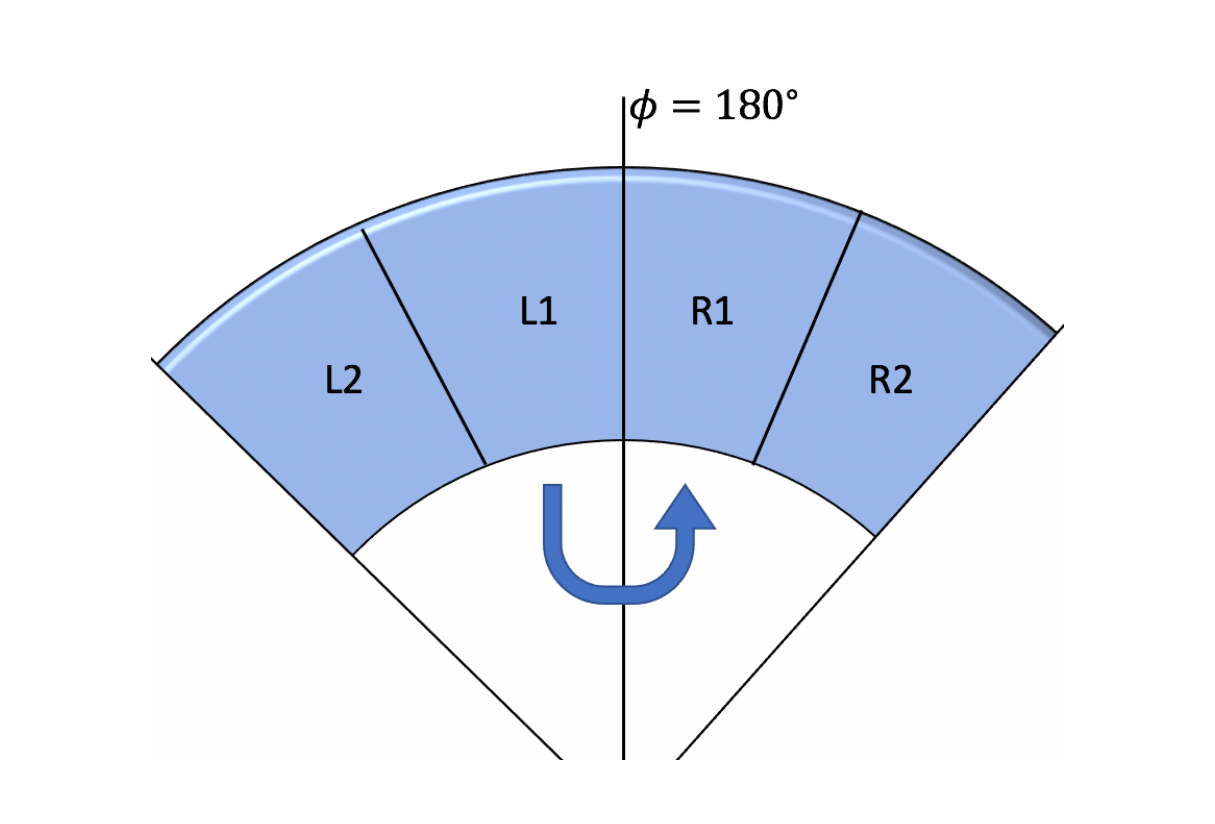}}
    \subfloat[]{\includegraphics[clip, trim=0.0cm 7.0cm 0.5cm 7cm, scale=0.4]{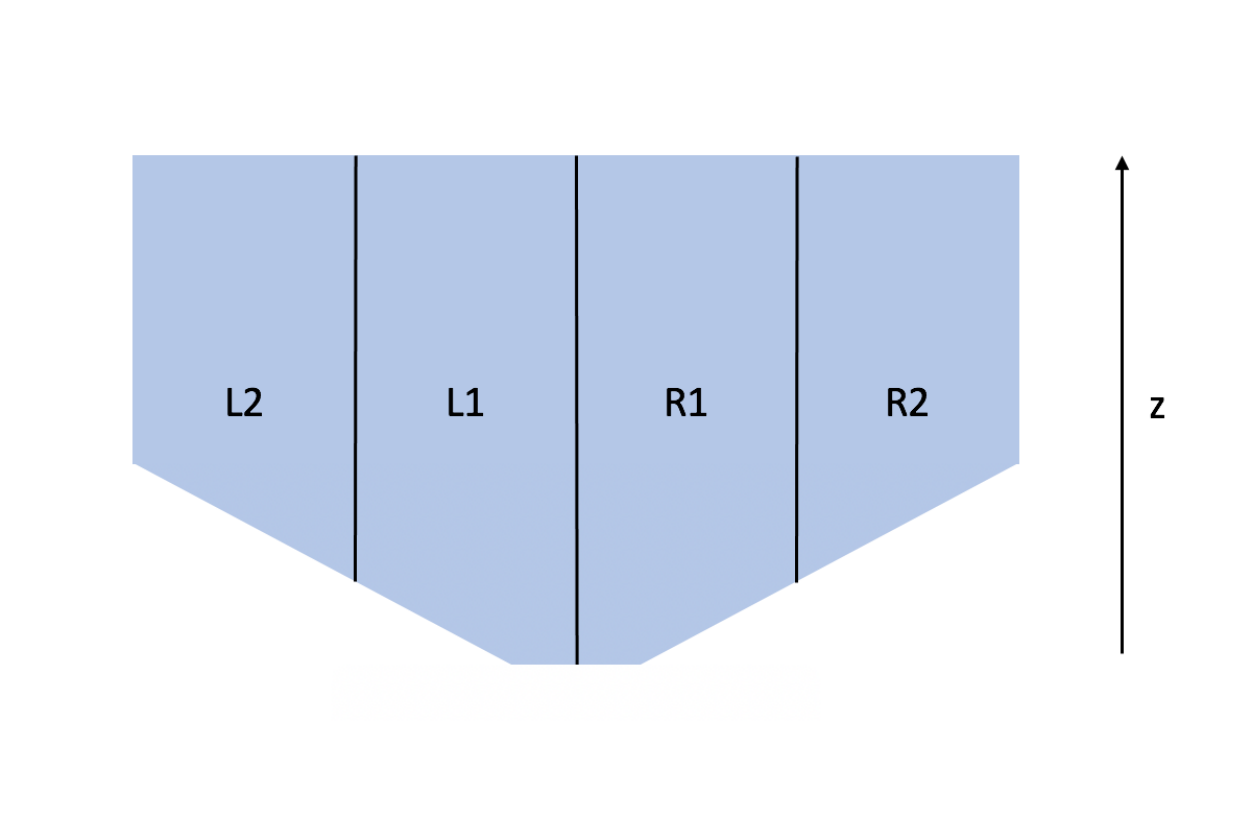}}
    \caption{ An illustration of the axial 
    symmetry we exploit in order to build a data vs. data 
    LS correlator. 
    (a) A schematic 
    of our data set projected onto the mid-plane as viewed from the North of the Galaxy.  
    Using the $\phi=180^{\circ}$ ray as the mirror line, the reflection of the data from L1 is compared with R1, while the data from L2 is compared with R2. 
    (b) The Northern half of the same model projected onto the $y-z$ plane at $x \approx -8$ kpc, the Sun's location, 
    as viewed from the Galactic center.  In this view of the rough schematic, the vertical structure of a reflection of L1 across the $\phi=180^{\circ}$ ray can be compared to the vertical structure of R1, as the structure examined in $z$ could change with $\phi$. 
    An analogous method is used for the reflection symmetry about $z = 0$. Here 
    we have chosen regions in $R$ and $\phi$ such that 
    $z_{\odot}$ can be neglected with little consequence,
    as we have noted in Fig.~\ref{fig:modelModel3}c.
    }
  \label{fig:symmetryExplainer}
  \end{center}
\end{figure}

In this paper we consider the
Gaia vs. Gaia 2PCF to probe the structure
associated with the broken symmetries observed in \citet{widrow2012galactoseismology,yanny2013stellar,ferguson2017milky,bennett2018vertical}, and \citet{GHY20},
though we suppose that diffuse structures such as dissolved or dissolving clusters or streams, if sufficiently large and massive, could potentially contribute to the 2PCF as well.  To this end, we compute the correlations 
in the vertical direction by comparing data from the North vs. a reflection (across $z=0$) of the data from the South and comparing data from the Right ($\phi < 180^{\circ}$) with a reflection (across $\phi = 180^{\circ}$) of data from the Left ($\phi > 180^{\circ}$).
These two comparisons test reflection and axial symmetry,  respectively.  Our selections in $R$ and $\phi$ in this study 
correspond to the case
described in Fig.~\ref{fig:modelModel3}c, and we  
set $z_\odot=0$ in this section. 
We illustrate the geometry of these selections in 
Fig.~\ref{fig:symmetryExplainer}.


Moreover, we are able to probe structure in the $x$ and $y$ directions as well, and we do so by computing the correlations in the $x$ and $y$ directions from comparing data on the Right of the $\phi = 180^{\circ}$ line with a reflection of the data on the Left of this line.  As our analysis is close to the $\phi = 180^{\circ}$ line, the $x$ and $y$ directions are proxies for the radial and azimuthal directions, respectively --- and we pick rectilinear
coordinates so that we can combine the computed correlations
across the sample. We compute the separation between
two stars as 
$|\hat{e}_Q\cdot \mathbf{x}_{12}|$ for $Q \in x,y,\, {\rm or}\, z$, 
as we define after Eq.~(\ref{x12defn}). 
These separations
are logged into the $RR$, $DD$, and $DR$ histograms in order to construct the LS correlator 
for several small regions of the Galaxy.  This ``Gaia vs. Gaia" method allows for the examination of structure in each coordinate direction independently, and it is completely independent of any theoretical model, though we do assume that our data set is free
from any artificial symmetry-breaking 
effects, which have been 
carefully assessed 
in Sec.~2.4 of \citet{HGY20}. 
This is in contrast to the model-dependent analysis of \citet{kamdar2021preliminaryclusteringindisk} which also 
utilized the full three-dimensional distance between 
two stars. Since our analysis is focused on the 
identification of symmetry-breaking effects, we 
also believe it to be particularly sensitive to 
non-steady-state effects. 

Overall, our method does require a large number of stars free of significant biases.  
As such, we employ the 11.7 million stars in the data set of \citet{HGY20} which has been selected to minimize the impact of faint-end incompleteness, crowded fields, extinction from dust, and artificial basis from  
the Gaia scan law. The selection satisfies 
$7 < R < 9$ kpc, $174^{\circ} < \phi < 186^{\circ}$,  $0.2 < |z| < 3.0$ kpc, $|b| > 30^{\circ}$, $14 < G < 18$ mag, and $0.5 < G_{\rm BP} - G_{\rm RP} < 2.5$ mag, to
yield 
an average uncertainty parallax of some $8.6\%$. 
With this data set, we are able to subdivide the 11.7 million stars into smaller wedges which still possess 
$\sim 10^5$ stars in a typical, small region close to the plane
and $\sim 50,000$ stars in the most limiting, high-$|z|$ cases, 
with an average completeness 
in excess of 99\% \citep{HGY20} when compared to number counts from the Hubble Space Telescope, as both dim stars and crowded fields are avoided \citep{arenou2018gaia}. 
This selection intersects the regions which contain the Gaia snail shell pattern \citep{antoja2018dynamically} and vertical waves \citep{widrow2012galactoseismology}, as well as the corrugation patterns noted in the 
simulation of 
\citet{blandhawthorn2021snail}. This 
also holds for
the more limited choices in $R$ and $\phi$ we 
make in our current study. 
It should also be noted that our study relies on comparisons between nominally symmetric portions of the Galaxy.  Therefore, we do not anticipate significant differences (say, North and South) in stellar crowding, parallax error, and other systematics between the portions of data being compared, in light of the elimination of scan-law patterns in our data set \citep{HGY20}.  Indeed, \citet{HGY20} explicitly quantifies
these systematic limitations.

Finally, we emphasize that 
our interest 
in small-scale structure and symmetry breaking has prompted us to examine the {\it Gaia} data in smaller regions. 
This subdivision of the data aided the computational efficiency of the study, 
as cross-correlations between various wedges of data were not computed.  The analysis was made quicker still by computing only one component of the pair-wise displacement.  Nonetheless, each small wedge of the Galaxy in our analysis possesses 
${\cal O}(10^5)$ stars (and thus ${\cal O}(10^{10})$ pairs), and so a \texttt{C++} program, optimized for speed, was written to analyze the large number of pairs.

\section{Analysis} \label{sec:Analysis}

We now turn to our Gaia vs. Gaia 2PCF analysis, 
employing the LS correlator of Eq.~(\ref{eq:x12_LS_MW}), considering first its vertical structure for 
different selections in $R, \phi$ and then its radial and azimuthal 
structure, studying the North and South separately. 
Since our LS correlator is not explicitly $D\leftrightarrow R$
symmetric, the choice of $D$ and $R$ can impact the final 
result, in principle. 
Thus in building the Gaia vs. Gaia 2PCF 
we assign the first data set (North) to ``$D$'' and the second
data set (South reflection) to ``$R$''. 
We ignore
the effect of $z_\odot$ throughout, so that we set $z_\odot=0$. 
As a result we consider selections within the region 
for which $R\in [7.6, 8.4]\,\rm kpc$, 
$\phi \in [178^\circ, 182^\circ]$, and  $|z| \in [0.2,3]\,\rm kpc$.

\begin{sidewaysfigure}
  \begin{center}
    \includegraphics[clip, trim=0.0cm 0.5cm 0.0cm 0.5cm, scale=0.655]{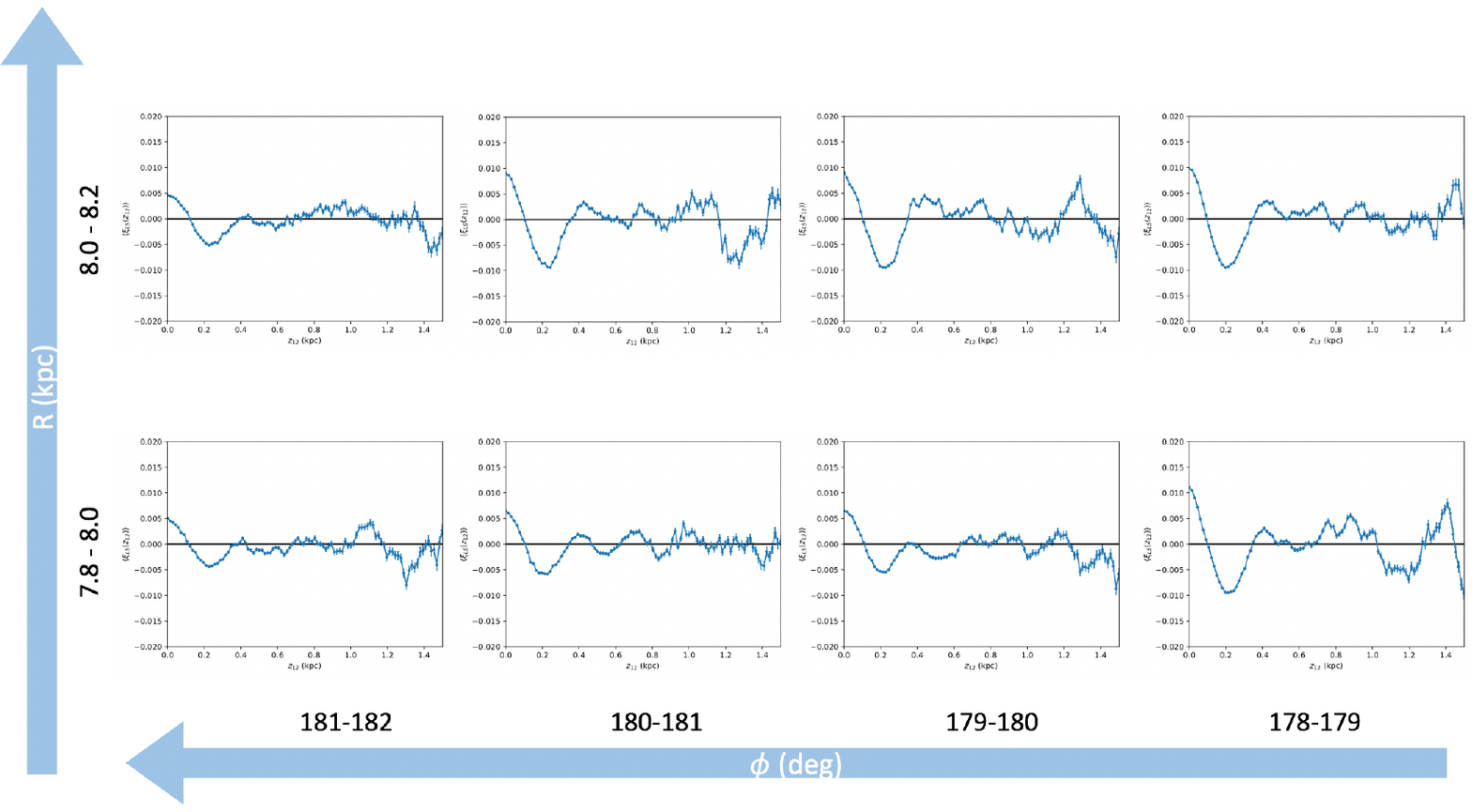}
    \caption{Vertical Structure probes:
    Several Gaia vs. Gaia, z2PCF 
    computations for various $R, \phi$ wedges, comparing the 
    North to the South.  The region pictured exhibits highly correlated stars, revealing vertical waves in a new light.  As the smallest, limiting length scale for this figure is about 40 pc, all of the larger wavelength features seen in the z2PCF
    would appear to be significant, 
    while the shorter wavelength features are more likely not.  Compare with Fig.\ref{fig:modelModel1}.
    }
  \label{fig:zPanel_NS}
  \end{center}
\end{sidewaysfigure}

\subsection{Vertical Structure} \label{subsec:VerticalStructure}
For the vertical 2PCF
analysis (z2PCF hereafter) we examine data selections
that form annular wedges in $R,\phi$, choosing a range of 
200 pc in $R$ and $1^{\circ}$ in $\phi$, with $z$ satisfying 
$0.2 < |z| < 2.0$ kpc. 
The separation distances are computed up to 1.5 kpc in separation, as the geometry fundamentally limits the number of pairs near the maximal 1.8 kpc of separation possible in the wedges, corresponding to stars at the maximal $|z| =$ 2.0 kpc and minimal $|z| =$ 0.2 kpc.  Each bin for the North vs. South analysis has a width of 15 pc, 
for a rough sampling size of $10^8$ pairs per bin, given the large
number of pairs possible in a sample of $10^5$ stars distributed over 100 bins, e.g.

\begin{figure}[h!]
  \begin{center}
    \hspace*{+0.8cm} 
    \subfloat[]{\includegraphics[scale=0.67]{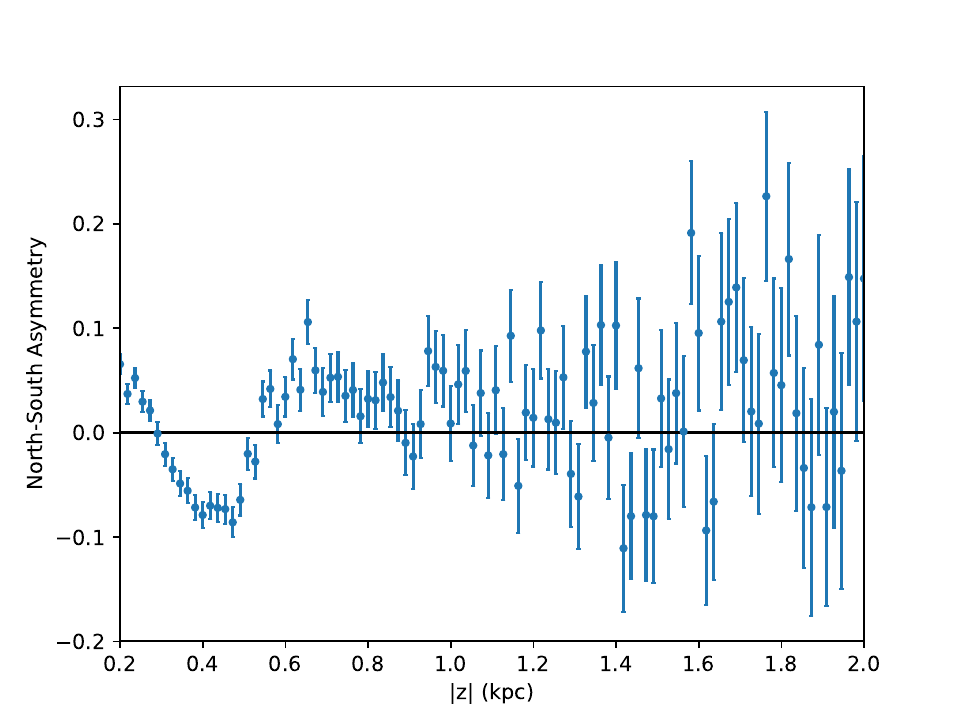}}
    
    \subfloat[]{\includegraphics[scale=0.67]{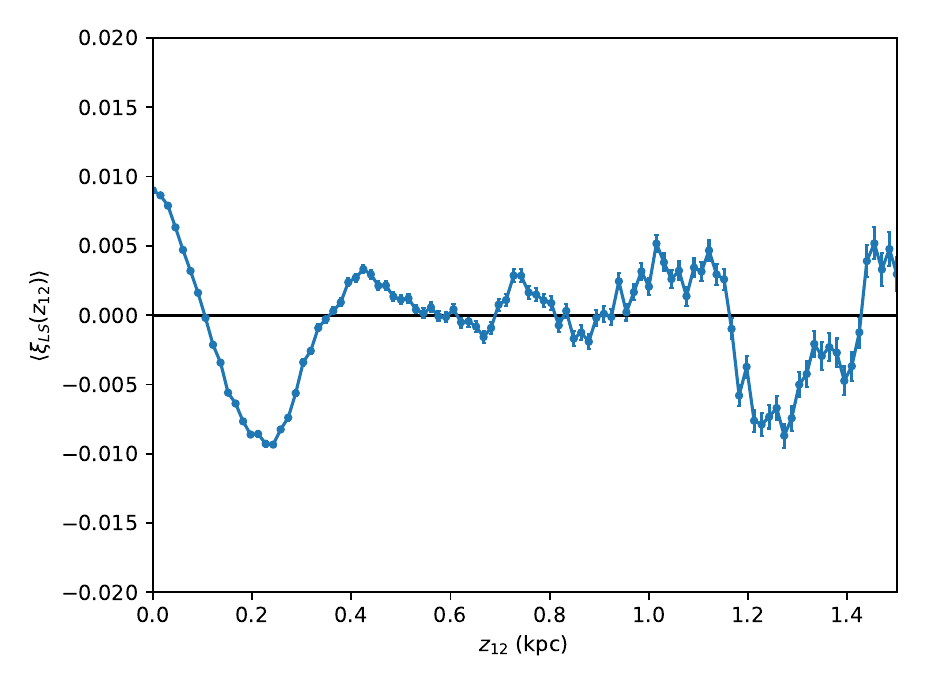}}
    \caption{
    A comparison of (a) the North-South asymmetry and (b) the Gaia-Gaia z2PCF, 
    North vs. South, in the region for which $R\in [8.0, 8.2]\,\rm kpc$, 
    $\phi \in [180^\circ, 181^\circ]$, and  $z\in [0.2,2]\,\rm kpc$. 
    The peaks and troughs in the asymmetry can be linked to some -- but seemingly not all -- of the peaks in the z2PCF. Namely, the crests of the wave as seen in the asymmetry at $|z| \approx 0.2$ and $|z| \approx 0.6$ kpc result in the peak near a vertical separation distance, of about $z_{12} = 0.4$ kpc in the 2PCF.  The peak-to-trough distances between either of the the first two peaks and the first trough are around 0.2 kpc, and this registers as a cross-correlation in the LS estimator and thus a trough in the 2PCF near $z_{12} = 0.2$ kpc.  
    The vertical asymmetry plotted in panel (a) has been centered at an asymmetry of zero by subtracting the error-weighted mean from the raw asymmetry to aid in the comparison of the two plots, as a gross offset towards the North or South will not affect the 2PCF.  
    The vertical structure comes into focus much more clearly in the 2PCF due to the ${\cal O}(N^2)$ statistics when compared to the asymmetry analysis for the same region, as the latter runs out of statistics very quickly at high $|z|$.  
    }
  \label{fig:asym_2pcf_cf}
  \end{center}
\end{figure}

As noted in earlier studies \citep{widrow2012galactoseismology,yanny2013stellar,ferguson2017milky,bennett2018vertical,GHY20} of the 
one-body density, i.e., 
the stellar number counts, North and South, 
significant vertical structure 
exists in the Galaxy near the Sun. 
As we demonstrate in Fig.~\ref{fig:zPanel_NS}, 
the z2PCF 
is highly correlated,
with an array of wave-like structures across the selected
regions.  That is, 
the particular wave-like pattern observed differs from wedge to wedge. 
Interestingly, the most significant differences  
appear at the highest separation distances,
i.e., at high $z_{12}$, whereas the peak-to-trough feature at low
$z_{12}$ appears to be universal, even if some variations in its strength 
appears across the sample, with it attaining greatest
significance for $\phi<180^\circ$. 
The wave pattern seen 
bears comparison to the vertical wave features observed in the one-body 
density 
we have previously 
noted 
but benefit from the ${\cal O}(N^2)$ statistics afforded by pair counting statistics, as 
opposed to 
the ${\cal O}(N)$ statistics of star counts. 
We compare the North-South asymmetry, defining it for $z>0$ as 
\begin{equation}
    {\cal A}(|z|) = \frac{ n(z) - n(-z)}{n(z) + n(-z)} \,,
\end{equation}
where $n(z)$ is the number of stars at $z$ within our selected $R,\phi$ sample, 
and the z2PCF in 
Fig.~\ref{fig:asym_2pcf_cf}.
We note that the sign of the North-South asymmetry indicates whether excess counts exist in the 
North or South, but the sign of the z2PCF -- and the 2PCF in general 
-- does not have that 
interpretation. Rather, as per Eq.~(\ref{eq:x12_LS_MW}), 
the sign of the 2PCF is determined by whether direct ($DD + RR$)
or cross ($DR$) correlations dominate. The ${\cal O}(N^2)$ statistics of the 2PCF 
give a more sensitive view of the correlations within the sample than the North-South asymmetry. 
Indeed, the 2PCF appears to have more finely resolved structures than the 
corresponding asymmetry results.
These differences -- especially at higher $z_{12}$ -- may speak 
to differences in the vertical waves across the plane, or perhaps to additional, as yet unappreciated effects. 

\begin{figure}[ht!]
  \begin{center}
    \subfloat[]{\includegraphics[scale=0.57]{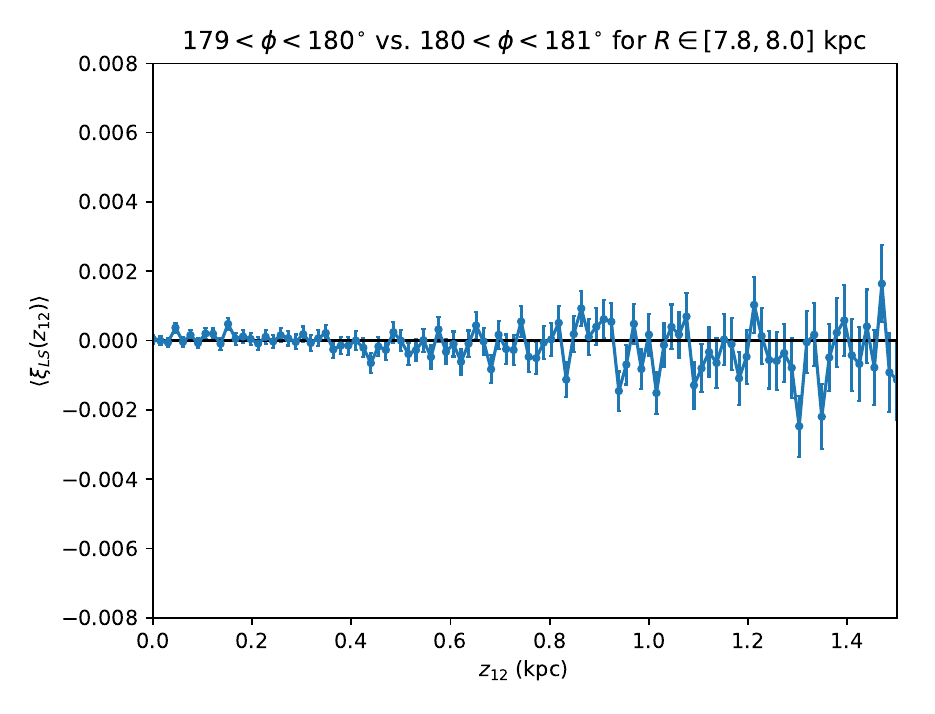}}
    \subfloat[]{\includegraphics[scale=0.57]{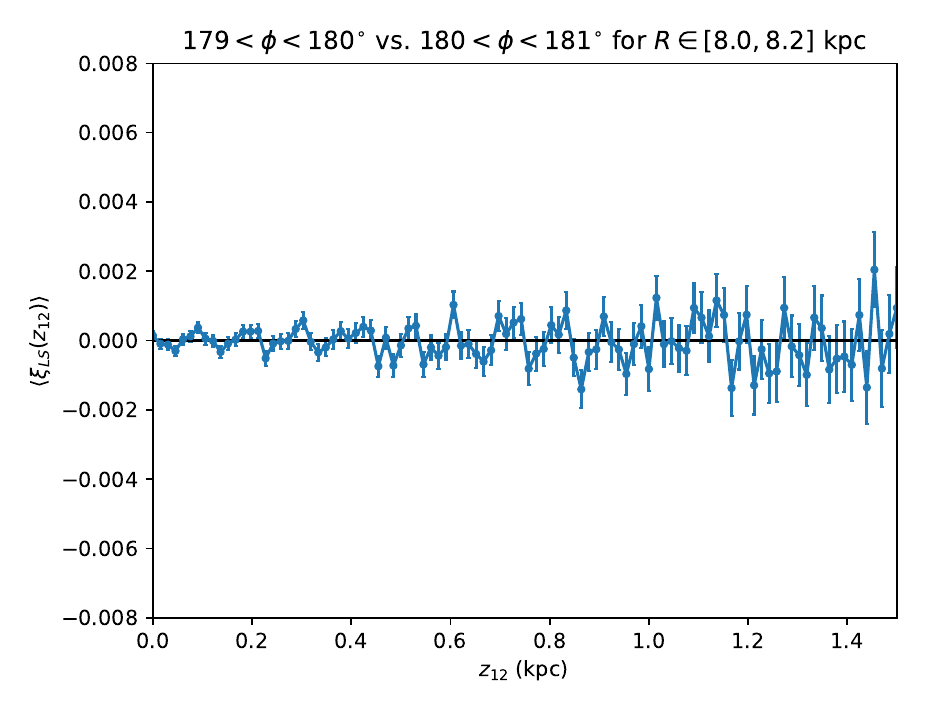}}

    \subfloat[]{\includegraphics[scale=0.57]{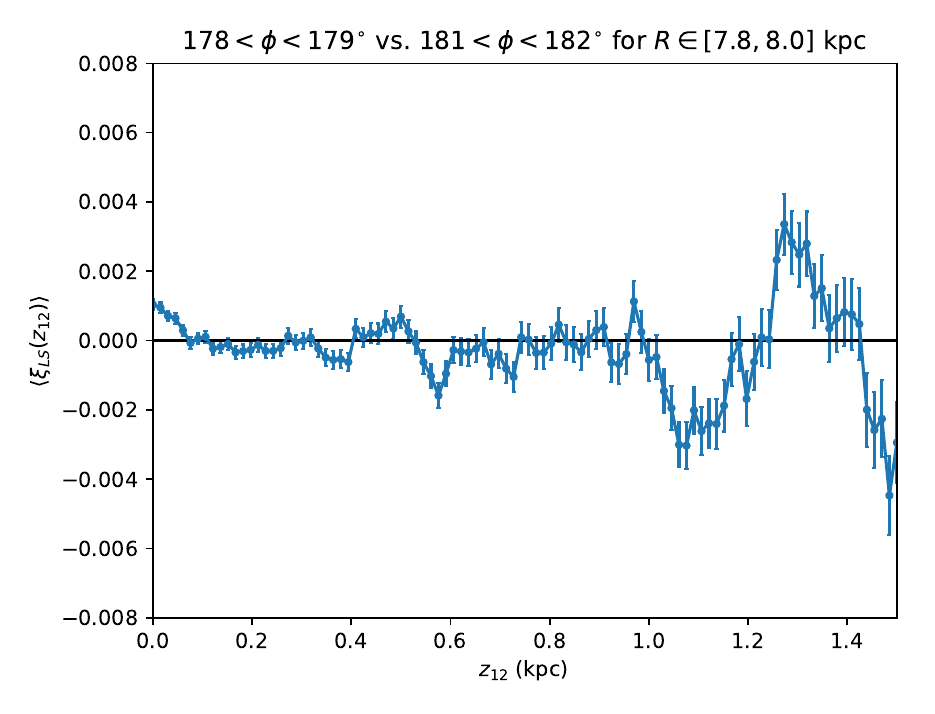}}
    \subfloat[]{\includegraphics[scale=0.57]{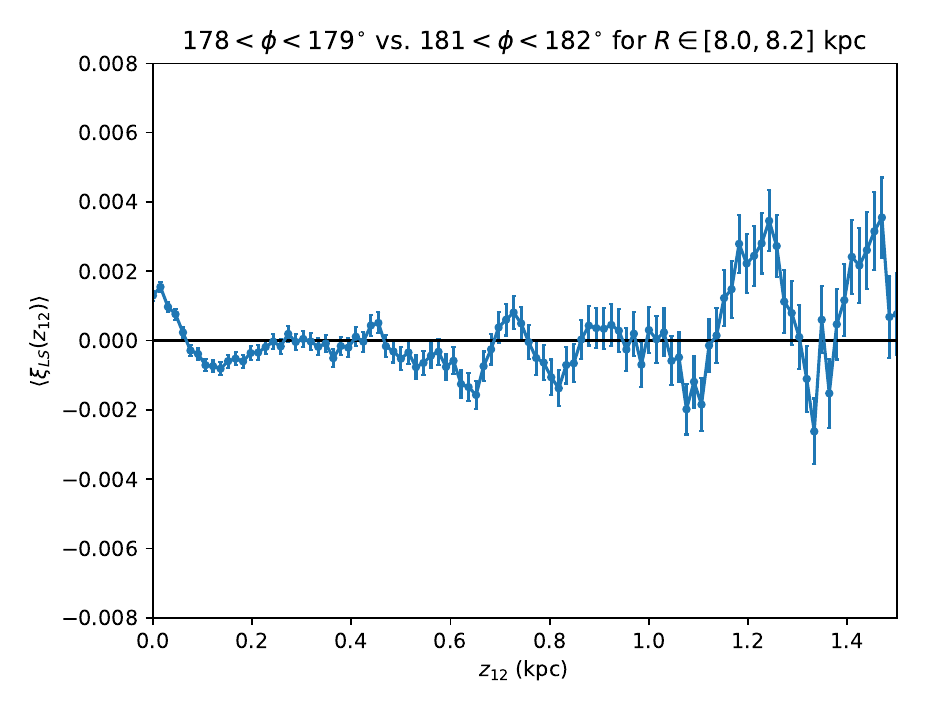}}
    \caption{Northern Hemisphere Azimuthal structure probes: 
    (a) The Left-Right z2PCF for $z > 0$ kpc, $7.8 < R < 8.0$ kpc, and $179^{\circ} < \phi < 180^{\circ}$ vs. $180^{\circ} < \phi < 181^{\circ}$.  
    (b) The Left-Right z2PCF for $z > 0$ kpc, $8.0 < R < 8.2$ kpc, and $179^{\circ} < \phi < 180^{\circ}$ vs. $180^{\circ} < \phi < 181^{\circ}$.  
    (c) The Left-Right z2PCF for $z > 0$ kpc, $7.8 < R < 8.0$ kpc, and $178^{\circ} < \phi < 179^{\circ}$ vs. $181^{\circ} < \phi < 182^{\circ}$.  
    (d) The Left-Right z2PCF for $z > 0$ kpc, $8.0 < R < 8.2$ kpc, and $178^{\circ} < \phi < 179^{\circ}$ vs. $181^{\circ} < \phi < 182^{\circ}$.  The smallest length scale to which we can probe is about 40 pc, indicating  significant structure at high values of $z_{12}$ in panels c and d 
    and a dearth of structure in panels a and b.
    }
  \label{fig:zPanel_LR_North}
  \end{center}
\end{figure}

\begin{figure}[ht!]
  \begin{center}
    \subfloat[]{\includegraphics[scale=0.57]{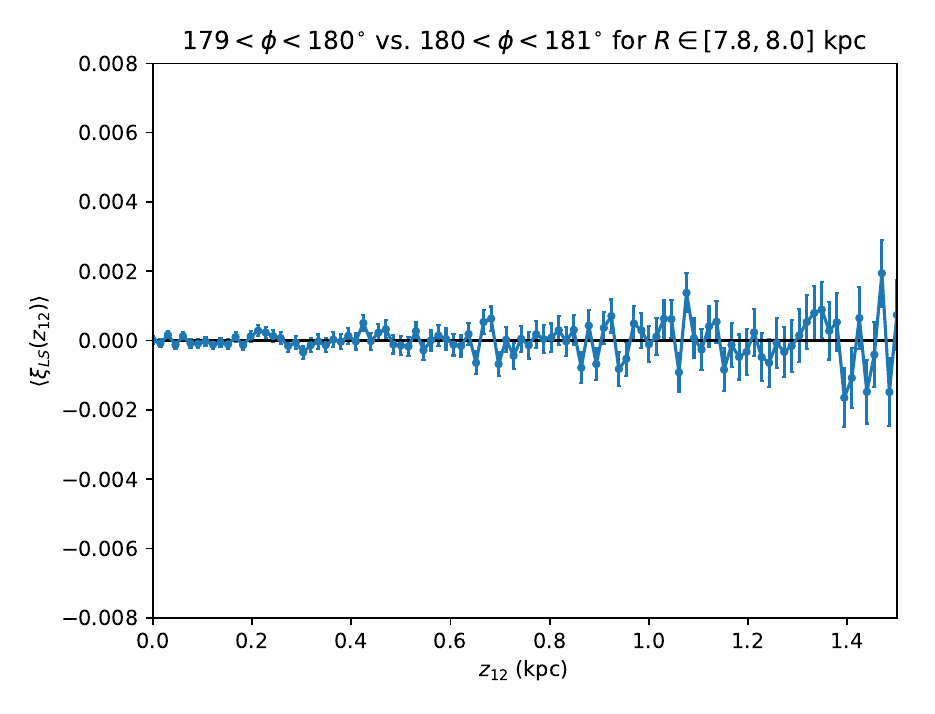}}
    \subfloat[]{\includegraphics[scale=0.57]{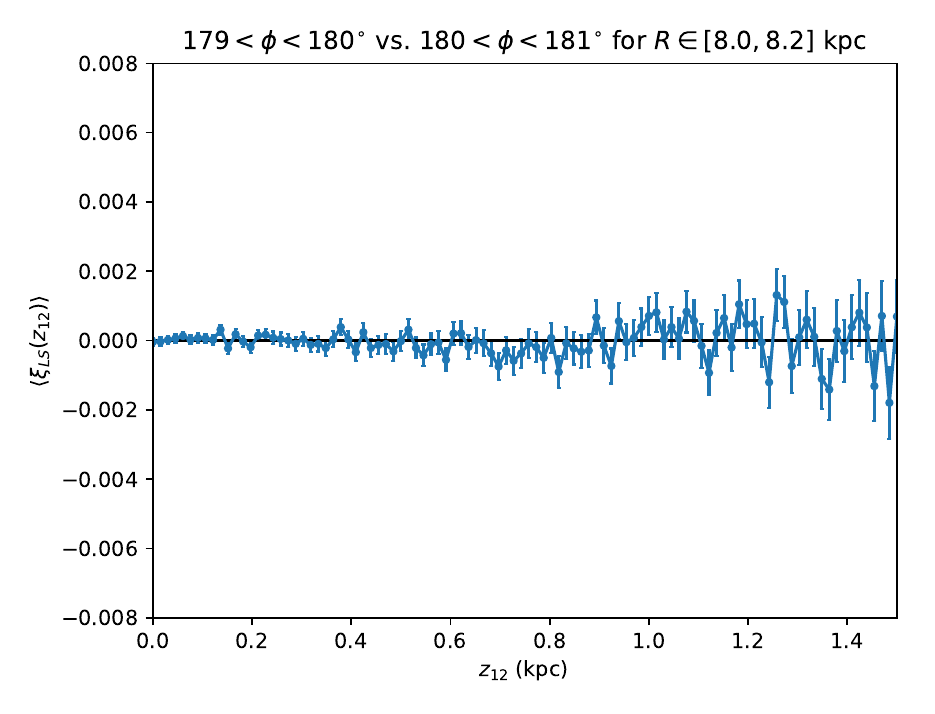}}

    \subfloat[]{\includegraphics[scale=0.57]{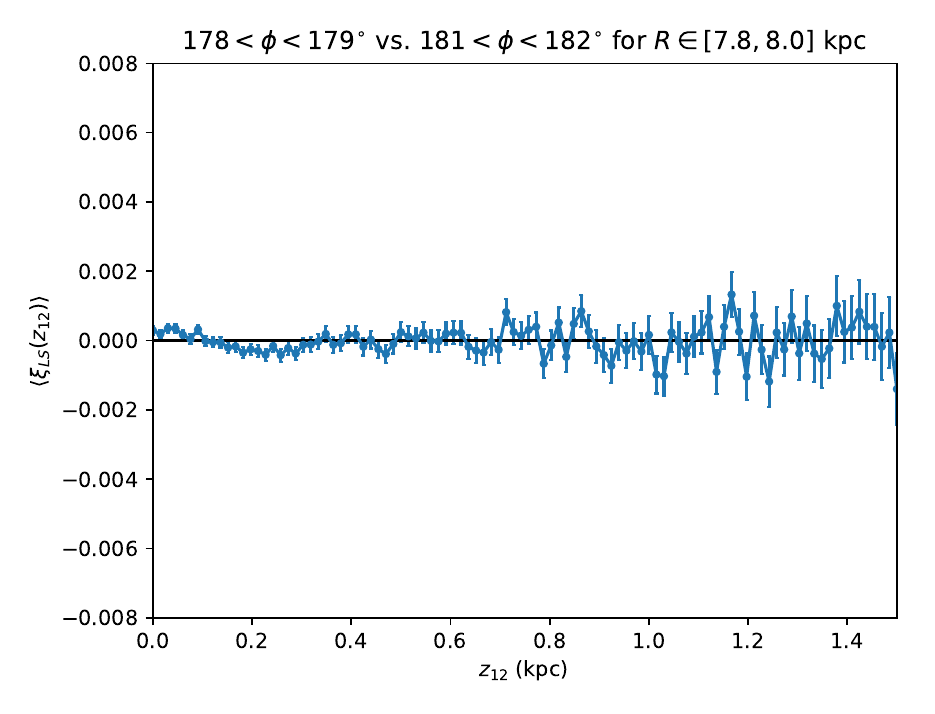}}
    \subfloat[]{\includegraphics[scale=0.57]{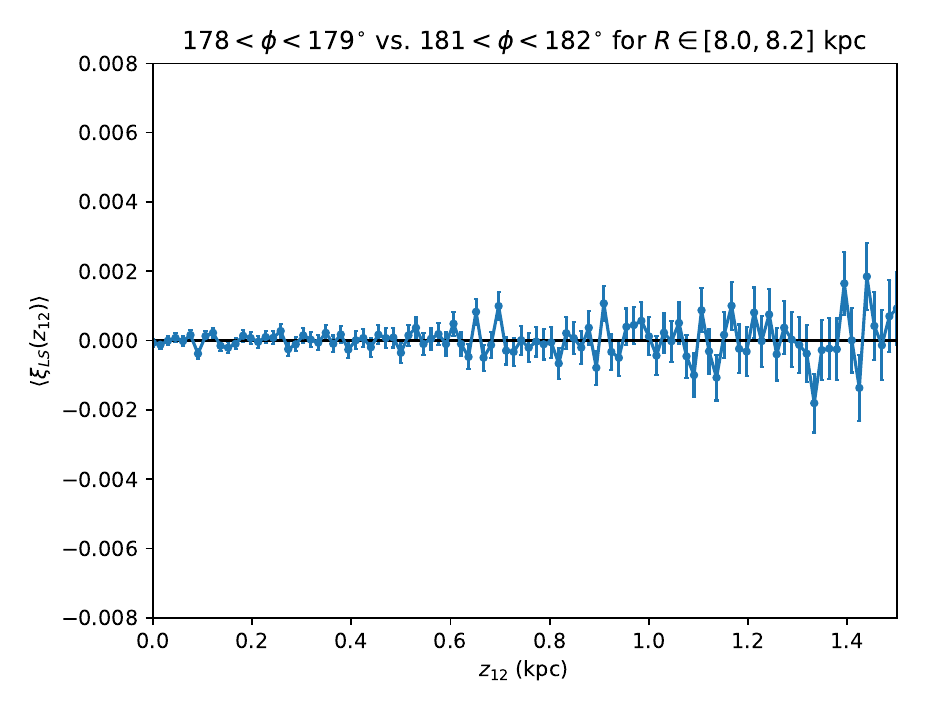}}
    \caption{Southern Hemisphere Azimuthal Structure probes:
    (a) The Left-Right z2PCF for $z < 0$ kpc, $7.8 < R < 8.0$ kpc, and $179^{\circ} < \phi < 180^{\circ}$ vs. $180^{\circ} < \phi < 181^{\circ}$.  
    (b) The Left-Right z2PCF for $z < 0$ kpc, $8.0 < R < 8.2$ kpc, and $179^{\circ} < \phi < 180^{\circ}$ vs. $180^{\circ} < \phi < 181^{\circ}$.  
    (c) The Left-Right z2PCF for $z < 0$ kpc, $7.8 < R < 8.0$ kpc, and $178^{\circ} < \phi < 179^{\circ}$ vs. $181^{\circ} < \phi < 182^{\circ}$.  
    (d) The Left-Right z2PCF for $z < 0$ kpc, $8.0 < R < 8.2$ kpc, and $178^{\circ} < \phi < 179^{\circ}$ vs. $181^{\circ} < \phi < 182^{\circ}$.  
    The smallest length scale to which we can probe is 40 pc, indicating a dearth of 
    differences with $\phi$ 
    in the vertical structure in the South.
    }
  \label{fig:zPanel_LR_South}
  \end{center}
\end{figure}

In addition to North-South differences, differences in the vertical structure from 
axial symmetry breaking 
can be assessed via a Left-Right comparison.  That is, a wedge of data on the left (say, $180^{\circ} < \phi < 181^{\circ}$) can be reflected across the $\phi = 180^{\circ}$ ray and compared against a wedge on the right (say, $179^{\circ} < \phi < 180^{\circ}$).  
Non-zero correlations from such a study speak to axial-symmetry breaking
in the vertical waves. 
Figures \ref{fig:zPanel_LR_North} and \ref{fig:zPanel_LR_South} illustrate these effects 
in the North and in the South, respectively.  
Different features 
are apparent.  First, structural variations appear for the waves in the Northern hemisphere, but hardly at all in the Southern hemisphere.  Additionally, the azimuthally adjacent wedges (panels a and b of Fig.~\ref{fig:zPanel_LR_North}) exhibit much less structure than wedges 
that are not azimuthally adjacent. This may just speak to larger effects in the 
$\phi< 179^\circ$ vs. $\phi> 181^\circ$ region, 
as also suggested in Fig.~\ref{fig:zPanel_NS}. 
Critically, however, the effect causing 
substantial differences in the wave structures with $\phi$ appear to be only 
significant in the North. 
Again, the largest differences between the various waves occur at higher values of $|z|$.

\begin{figure}[ht!]
  \begin{center}
    \subfloat[]{\includegraphics[scale=0.5]{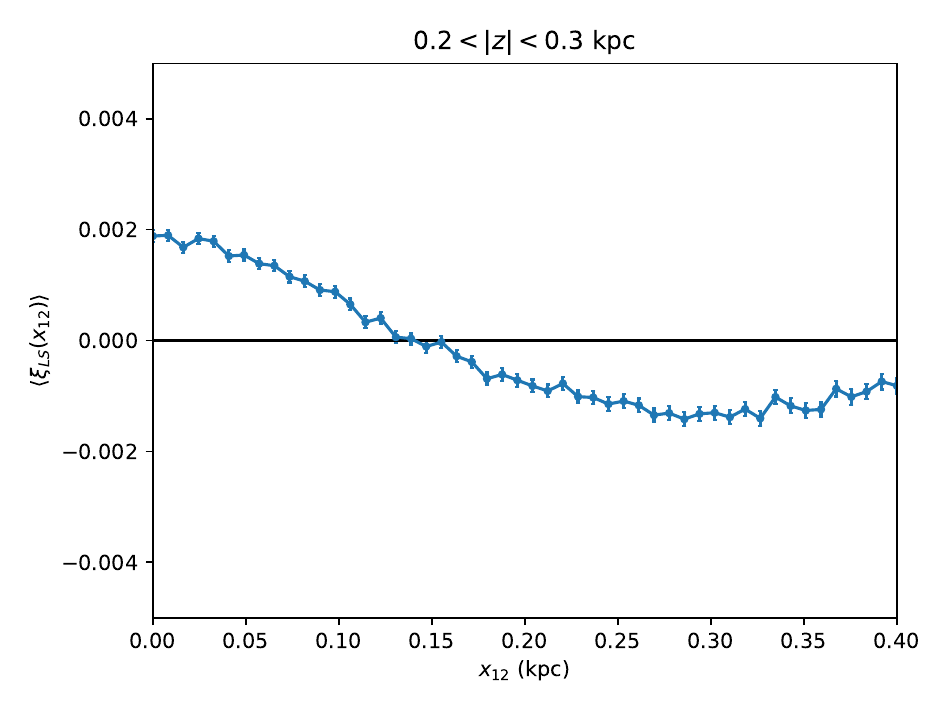}}
    \subfloat[]{\includegraphics[scale=0.5]{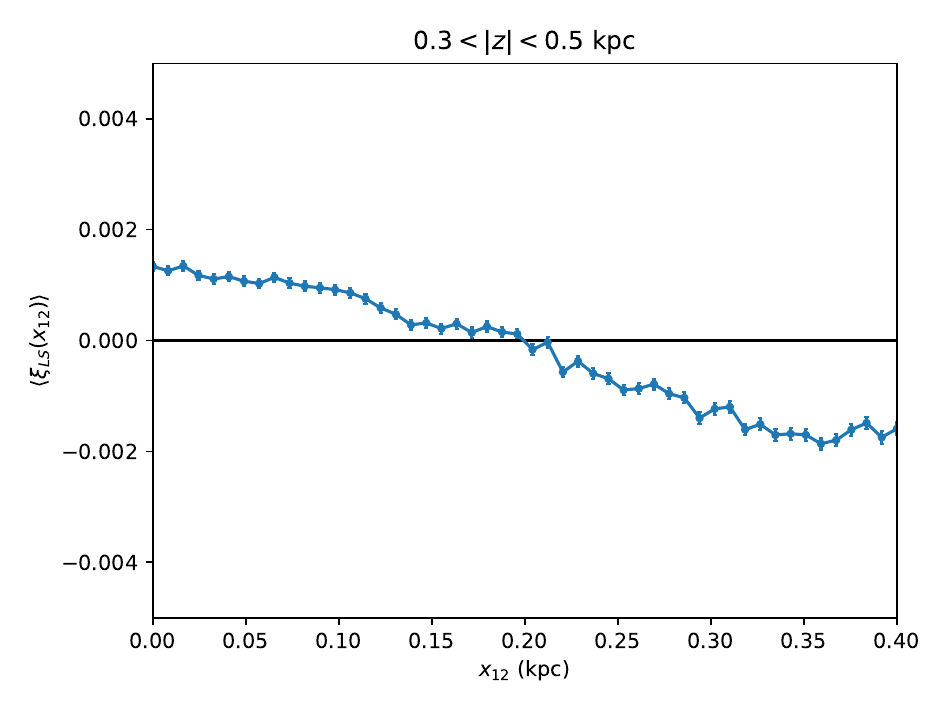}}

    \subfloat[]{\includegraphics[scale=0.5]{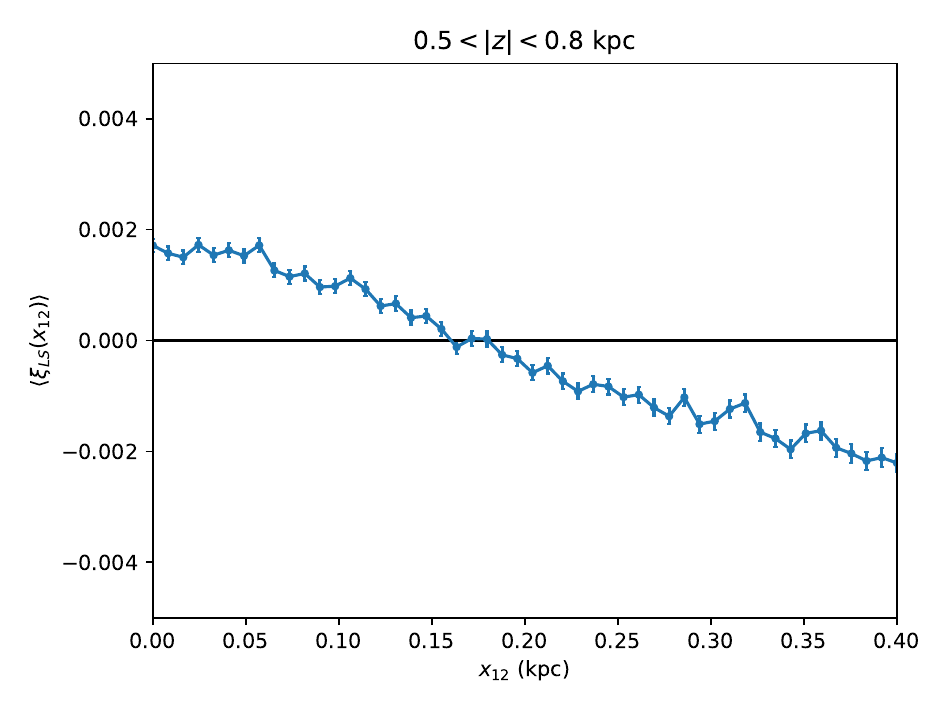}}
    \subfloat[]{\includegraphics[scale=0.5]{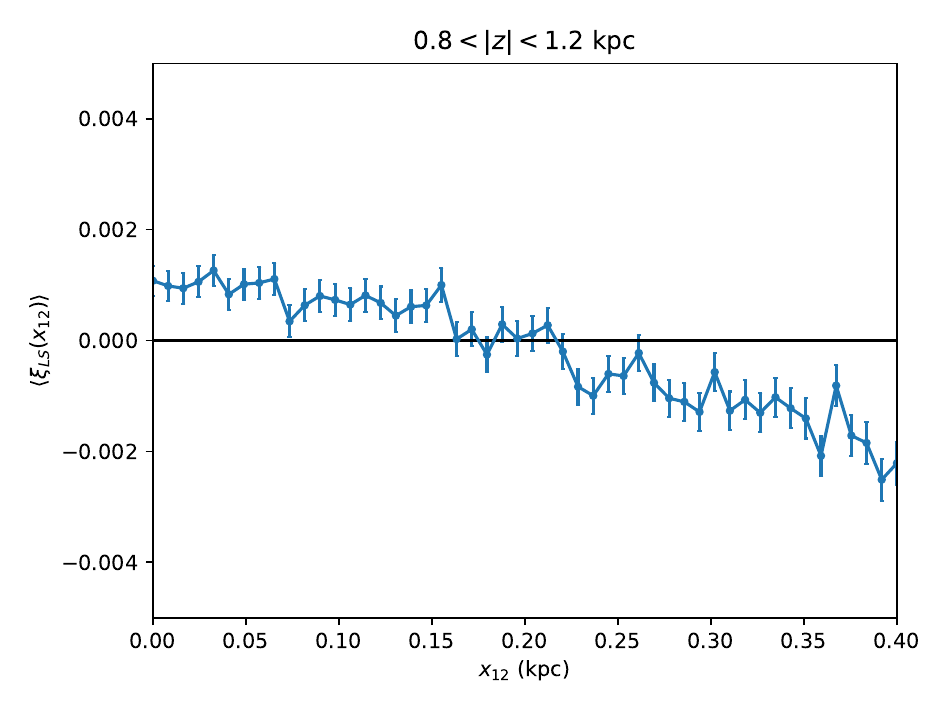}}
    
    \subfloat[]{\includegraphics[scale=0.5]{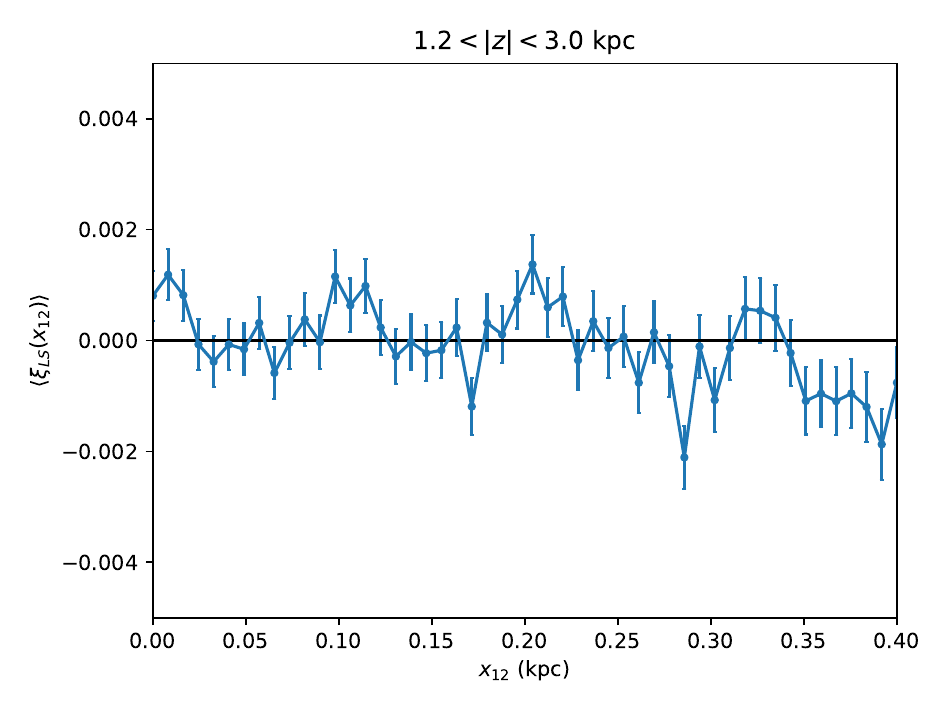}}
    \caption{Northern Hemisphere Radial structure probes:
    Left ($180^{\circ} < \phi < 182^{\circ}$) vs. Right ($178^{\circ} < \phi < 180^{\circ}$) comparisons of structure in the $x$-direction for the Northern hemisphere, with $7.6 < R < 8.4$ kpc, and for various slices of $|z|$.  (a) $0.2 < |z| < 0.3$ kpc, (b) $0.3 < |z| < 0.5$ kpc, (c) $0.5 < |z| < 0.8$ kpc, (d) $0.8 < |z| < 1.2$ kpc, (e) $1.2 < |z| < 3.0$ kpc.  The smallest length scale to which we can probe is 20 pc at lower values of $z$, and 30 pc at the highest range of $z$.
    }
  \label{fig:xPanel_LR_North}
  \end{center}
\end{figure}

\begin{figure}[ht!]
  \begin{center}
    \subfloat[]{\includegraphics[scale=0.5]{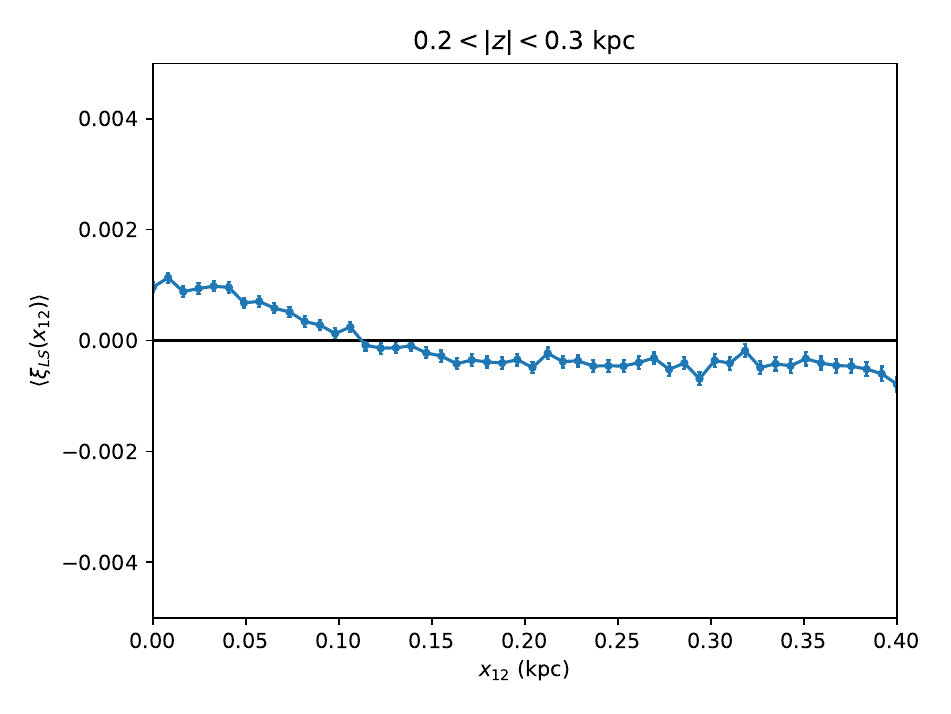}}
    \subfloat[]{\includegraphics[scale=0.5]{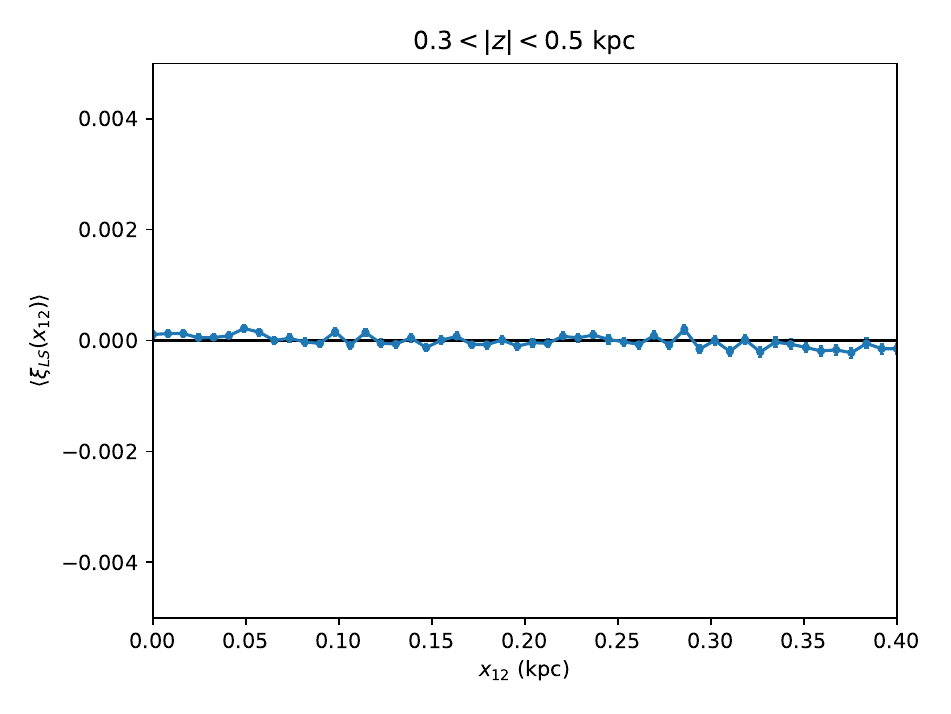}}

    \subfloat[]{\includegraphics[scale=0.5]{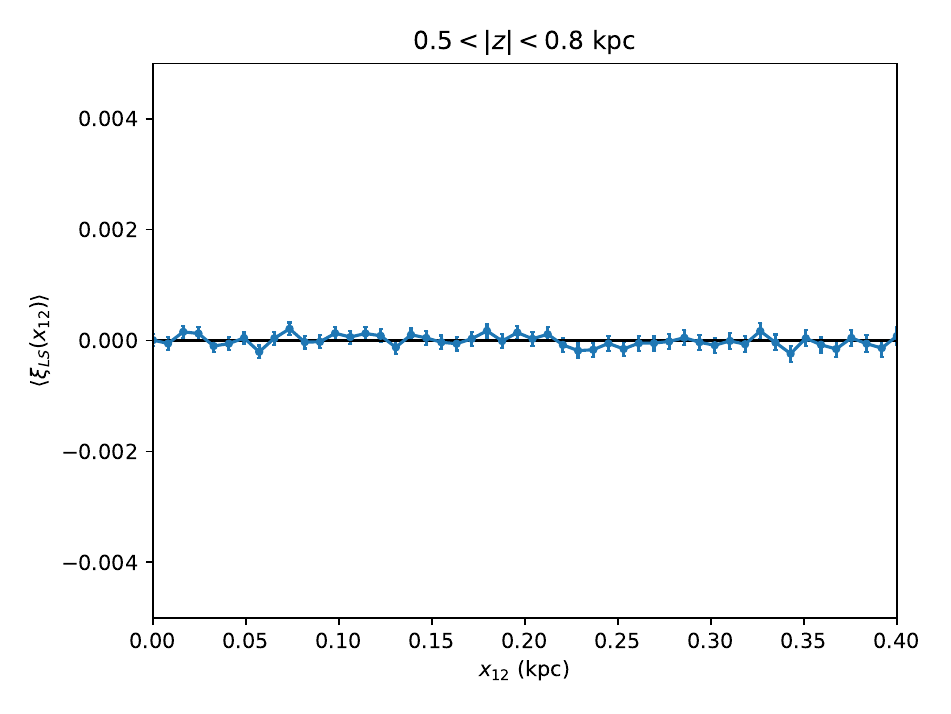}}
    \subfloat[]{\includegraphics[scale=0.5]{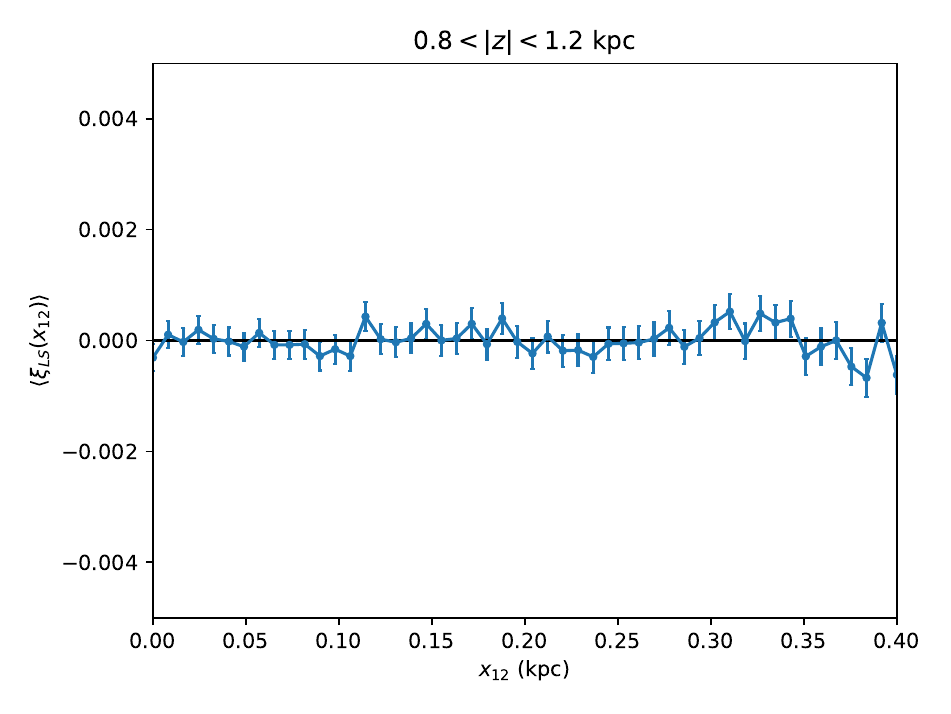}}
    
    \subfloat[]{\includegraphics[scale=0.5]{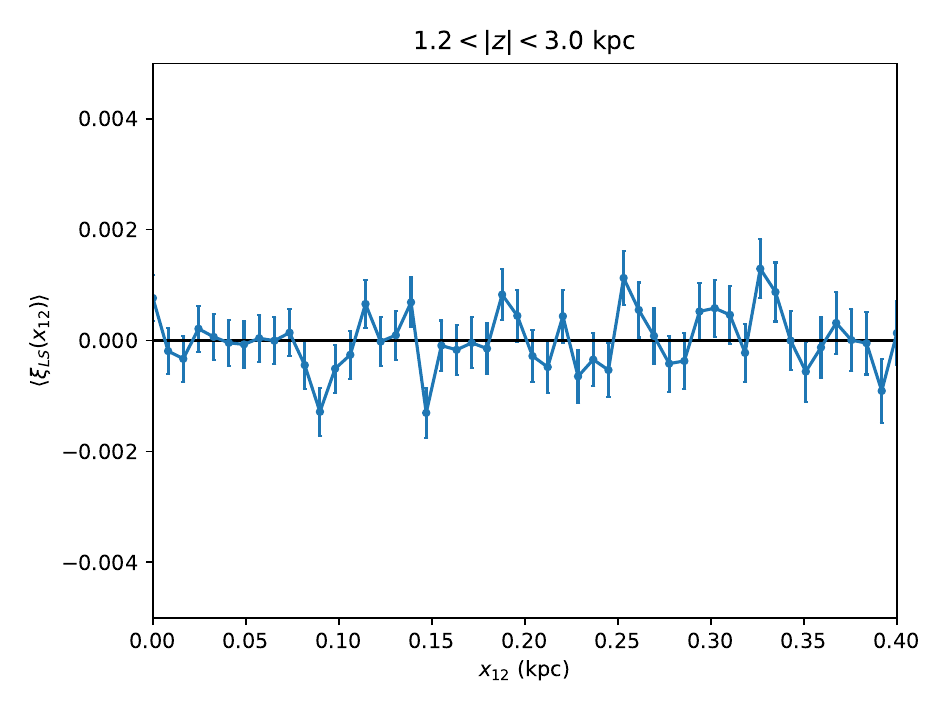}}
    \caption{Southern Hemisphere Radial Structure probes:
    Left ($180^{\circ} < \phi < 182^{\circ}$) vs. Right ($178^{\circ} < \phi < 180^{\circ}$) comparisons of structure in the $x$-direction for the Southern hemisphere, with $7.6 < R < 8.4$ kpc, and for various slices of $|z|$.  (a) $0.2 < |z| < 0.3$ kpc, (b) $0.3 < |z| < 0.5$ kpc, (c) $0.5 < |z| < 0.8$ kpc, (d) $0.8 < |z| < 1.2$ kpc, (e) $1.2 < |z| < 3.0$ kpc.  The smallest length scale to which we can probe is 20 pc at lower values of $z$, and 30 pc at the highest range of $z$.
    }
  \label{fig:xPanel_LR_South}
  \end{center}
\end{figure}

\begin{figure}[ht!]
  \begin{center}
    \subfloat[]{\includegraphics[scale=0.5]{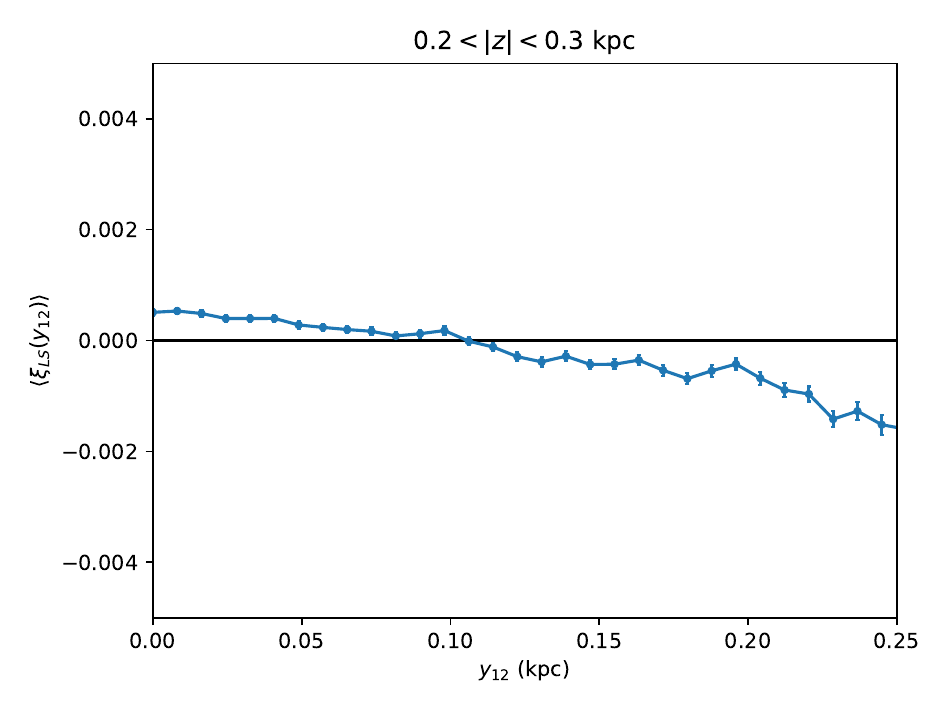}}
    \subfloat[]{\includegraphics[scale=0.5]{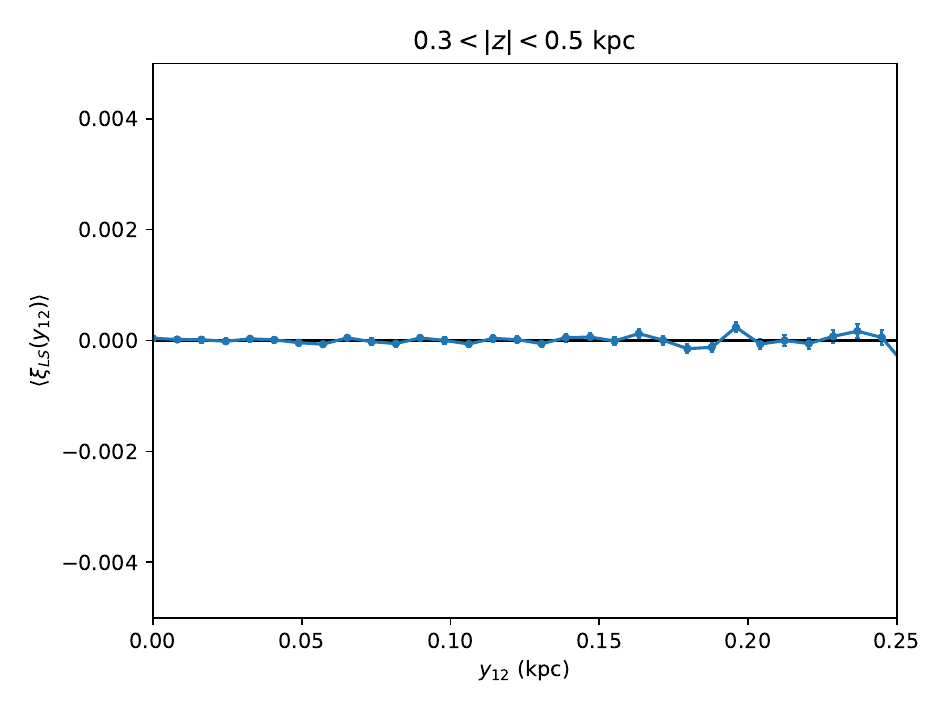}}

    \subfloat[]{\includegraphics[scale=0.5]{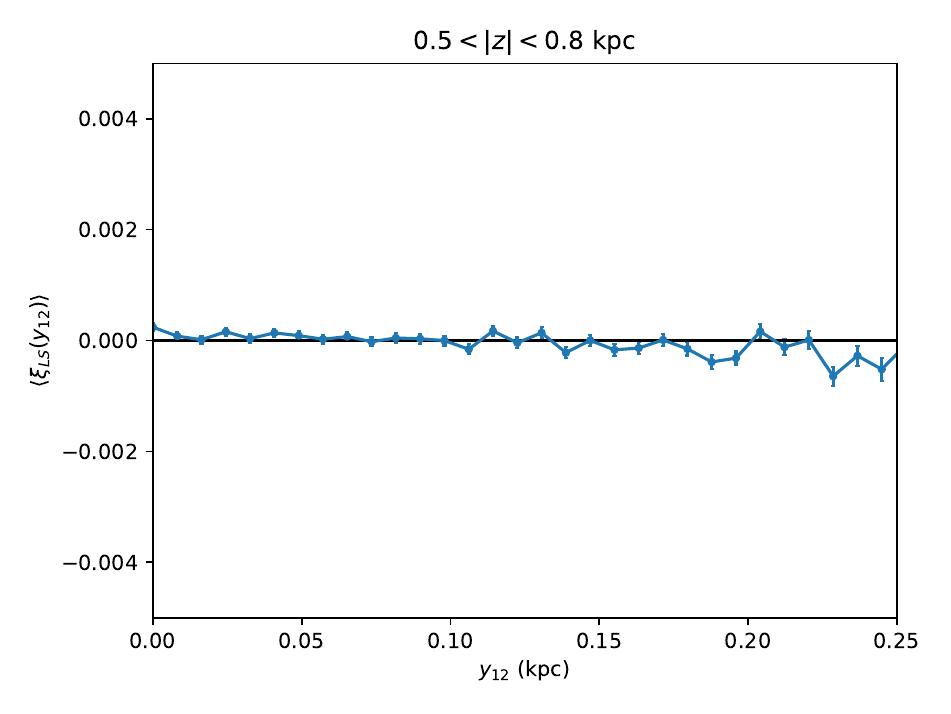}}
    \subfloat[]{\includegraphics[scale=0.5]{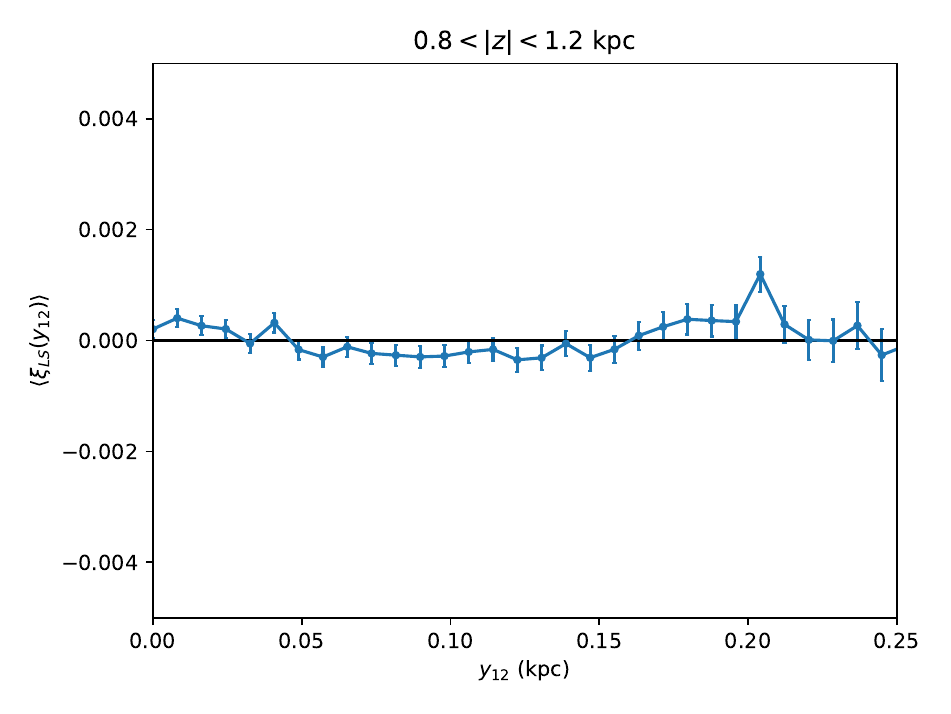}}
    
    \subfloat[]{\includegraphics[scale=0.5]{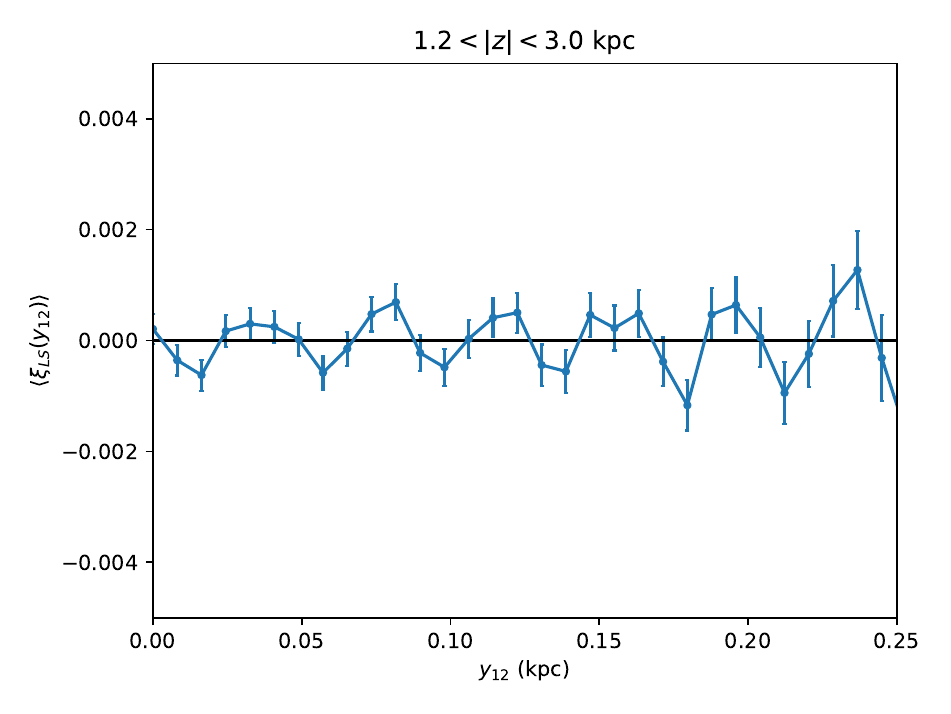}}
    \caption{
    Left ($180^{\circ} < \phi < 182^{\circ}$) vs. Right ($178^{\circ} < \phi < 180^{\circ}$) comparisons of structure in the $y$-direction for the Northern hemisphere, with $7.6 < R < 8.4$ kpc, and for various slices of $|z|$.  (a) $0.2 < |z| < 0.3$ kpc, (b) $0.3 < |z| < 0.5$ kpc, (c) $0.5 < |z| < 0.8$ kpc, (d) $0.8 < |z| < 1.2$ kpc, (e) $1.2 < |z| < 3.0$ kpc.  The smallest significant length scale is 10 pc for this figure.
    }
  \label{fig:yPanel_LR_North}
  \end{center}
\end{figure}

\begin{figure}[ht!]
  \begin{center}
    \subfloat[]{\includegraphics[scale=0.5]{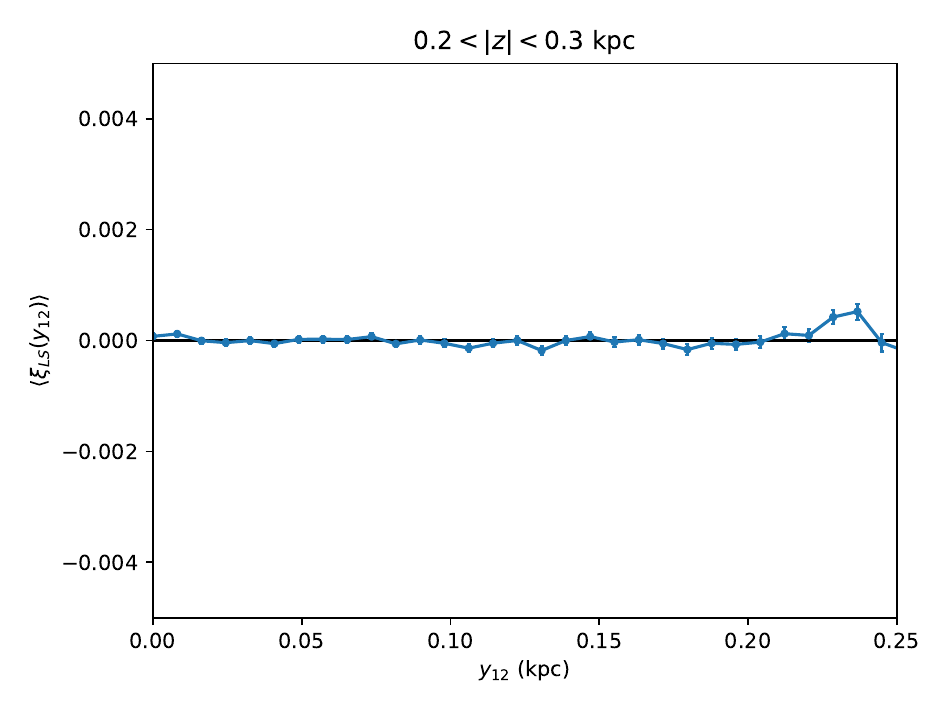}}
    \subfloat[]{\includegraphics[scale=0.5]{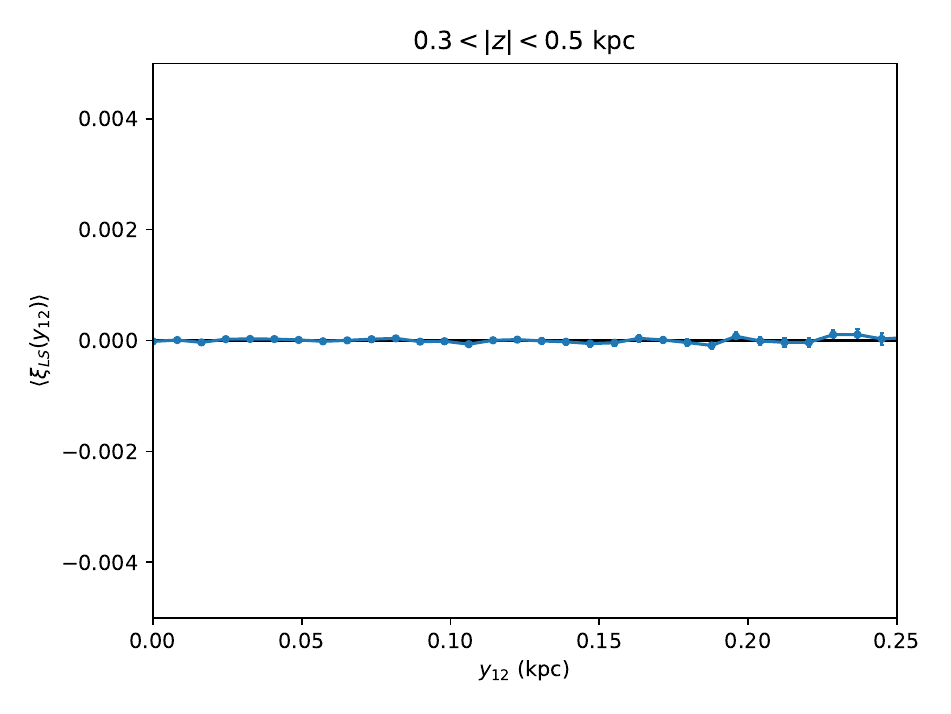}}

    \subfloat[]{\includegraphics[scale=0.5]{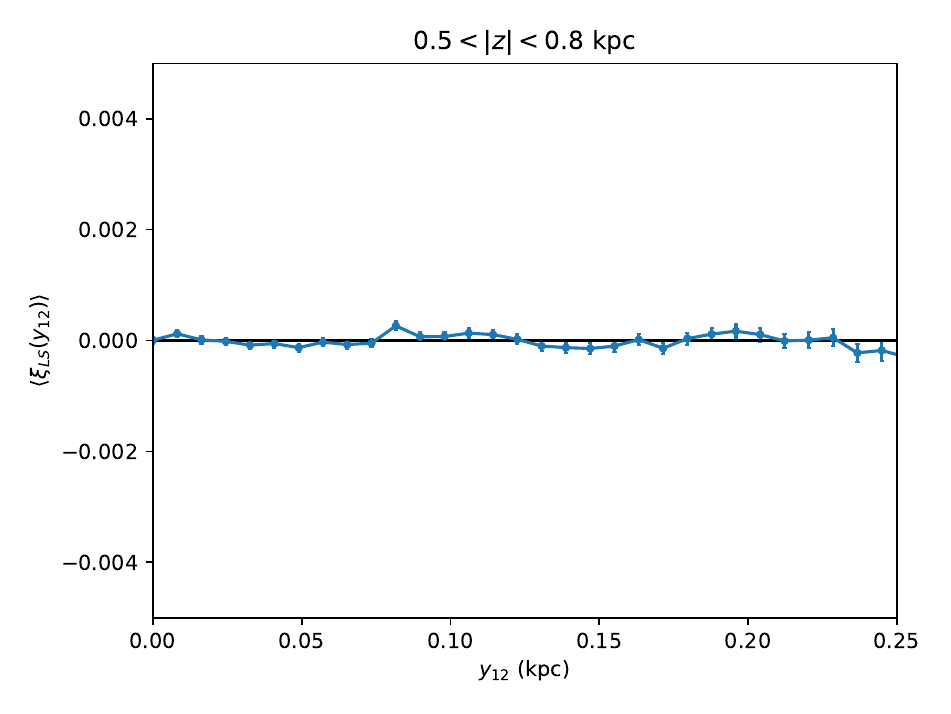}}
    \subfloat[]{\includegraphics[scale=0.5]{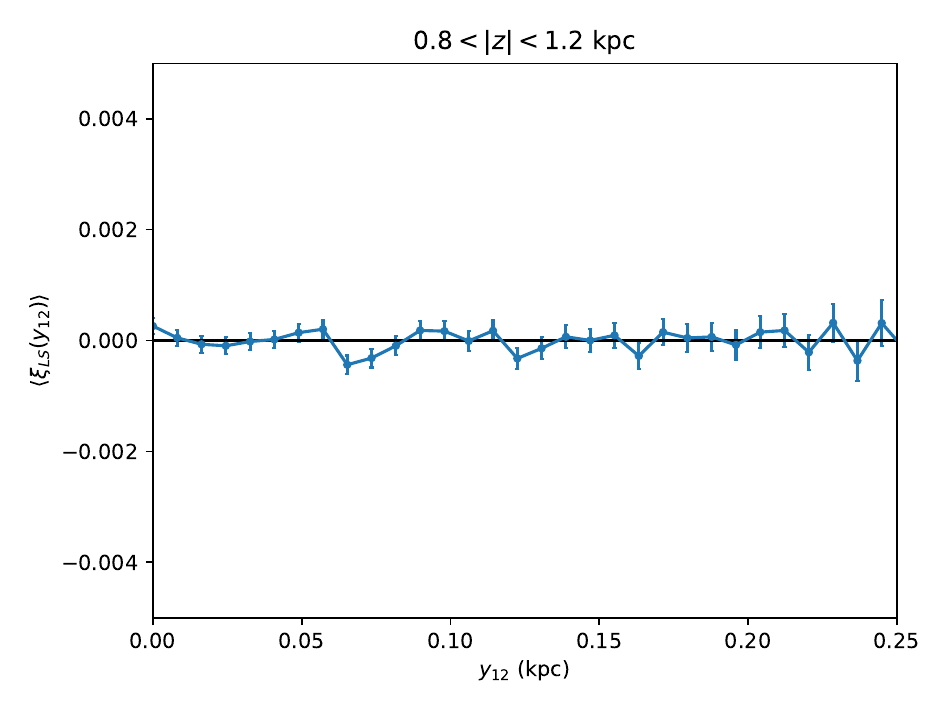}}
    
    \subfloat[]{\includegraphics[scale=0.5]{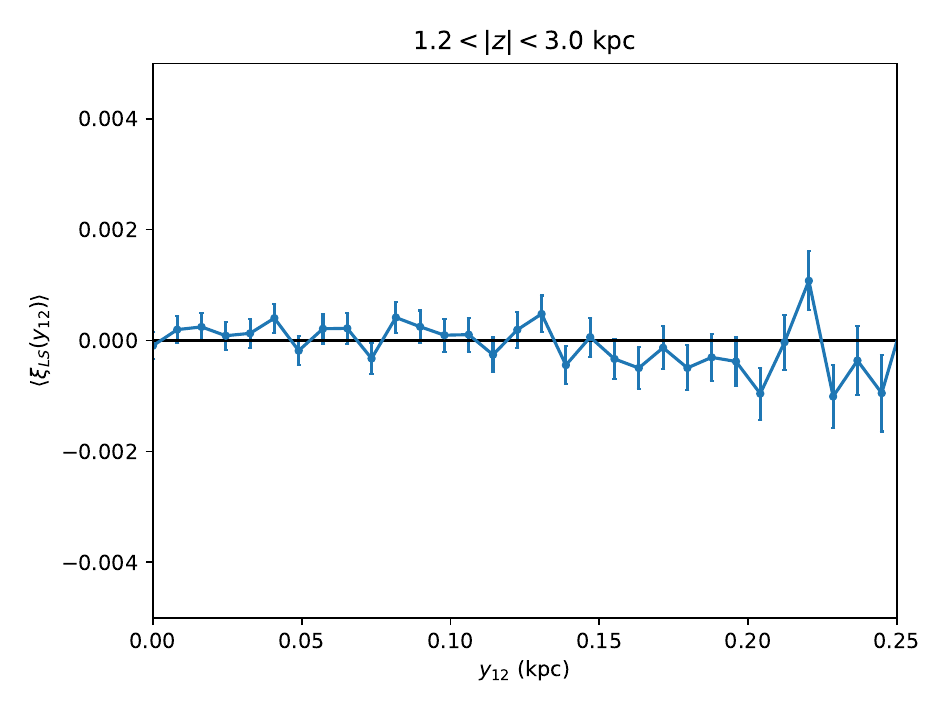}}
    \caption{
    Left ($180^{\circ} < \phi < 182^{\circ}$) vs. Right ($178^{\circ} < \phi < 180^{\circ}$) comparisons of structure in the $y$-direction for the Southern hemisphere, with $7.6 < R < 8.4$ kpc, and for various slices of $|z|$.  (a) $0.2 < |z| < 0.3$ kpc, (b) $0.3 < |z| < 0.5$ kpc, (c) $0.5 < |z| < 0.8$ kpc, (d) $0.8 < |z| < 1.2$ kpc, (e) $1.2 < |z| < 3.0$ kpc.  The smallest significant length scale is 10 pc for this figure.
    }
  \label{fig:yPanel_LR_South}
  \end{center}
\end{figure}

\subsection{Radial and Azimuthal Structure} \label{subsec:R_Phi_Structure}
We have seen how the vertical structure of the Galaxy can change substantially with 
Galactic radius and azimuthal angle, 
so that it is natural to ask if in-plane correlations exist as well.  Indeed, the Gaia 
phase-space spiral
\citep{antoja2018dynamically}, the observed 
axial-symmetry breaking in stars at heights well away from the Galactic
mid-plane \citep{GHY20, HGY20}, the axial differences in vertical structure noted in 
\citet{ferguson2017milky} and in this paper, 
as well as the complex corrugation patterns suggested by 
\citet{blandhawthorn2021snail} 
all point to the tangible possibility of non-zero radial and azimuthal correlations. 
Thus, we examine radial and azimuthal structure in the 2PCF in $x_{12}$ and $y_{12}$,
which we denote as x2PCF and y2PCF, respectively. 
We note that 
$x$ and $y$ act as proxies for $R$ and $\phi$ near the anti-Center line, as
we consider here. 
For the x2PCF and y2PCF 
analyses we once again examine annular wedges of data, 
this time 
with $7.6 < R < 8.4$ kpc and azimuthal widths of $2^{\circ}$ in $\phi$, 
and for various selections in $z$. 
 The separation distances are computed up to 0.4 kpc in $x_{12}$
 and to 0.25 kpc in $y_{12}$, 
 as the geometry 
 limits the number of pairs beyond these scales.  Each bin for the $x$ and $y$ analyses has a width of 8 pc.

Left-Right structural differences in the $x$-direction are shown in Fig.~\ref{fig:xPanel_LR_North} for various $z$ slices in the Northern hemisphere, and
the waves appear to become shorter in wavelength as 
$z$ increases
from panel (a) through panel (e).  Interestingly, the stars in the North are 
correlated in the $x$-direction, especially 
for $|z|< 1.2\,\rm kpc$.  The same is not true for the South, however, as depicted in Fig.~\ref{fig:xPanel_LR_South}.  While both the North and the South share some structural similarities at the lowest $|z|$ values, the South lacks the structure seen in the North in the region with $0.3 < |z| < 1.2$ kpc, 
as shown in panels b-d. The preponderance of $x$-direction 
structure in the North and the lack thereof 
in the South is also consistent with our vertical studies. 
Indeed, the vertical waves we observe exhibit marked Left-Right structural differences in the North (Fig.~\ref{fig:zPanel_LR_North}), but not the South (Fig.~\ref{fig:zPanel_LR_South}).  It appears that axial-symmetry breaking combines
with the vertical waves to 
to create a corrugation pattern of some kind, even beyond the region of the Gaia snail \citep{antoja2018dynamically}, though this appears to be restricted to the North. 

Additionally, slight hints of $y$-direction structure exist in the North (Fig.~\ref{fig:yPanel_LR_North}), but not the South (Fig.~\ref{fig:yPanel_LR_South}).  While not as significant as the $x$-direction structures we have just noted, 
the stars nonetheless appear to be correlated.  Moreover, a tantalizing, wave-like structure appears at the highest $|z|$ values in the North, perhaps hinting at a new
structure: an azimuthal wave of some sort.  The wavelength of this wave-like correlation pattern is about 40-50 pc.

We now turn to the consideration of the possible origins of the effects we have observed.

\section{Theoretical Origins}
\label{sec:origin}

The intricate phase-space correlations presented by 
the observed 
vertical asymmetries~\citep{widrow2012galactoseismology,yanny2013stellar,bennett2018vertical,Thulasidharan2021evidencevertical} and the existence of various snail-like correlations in 
position and velocity space~\citep{antoja2018dynamically,blandhawthorn2021snail,Gandhi2022snailacrossscales}
in the stars within roughly $1-2\,\rm kpc$ of the Sun 
have already prompted much discussion. In this section we revisit these suggestions 
and consider if any of them can explain the particular features we
have found. 
The symmetry-based 2PCF framework we have developed in this paper 
allows us to study the intricate three-dimensional nature of 
stellar 
correlations in position space, and a striking outcome of our study
is that the $R$ and $\phi$ correlations in the North are much richer than 
in the South. We think that the various phase-space correlations 
we have found and which have been noted 
shed light on the merger history of the Milky Way ---
and note, for context, that mergers have been invoked, variously, 
to explain the origin of the spiral arms~\citep{purcell2011sagittarius} and 
of the thick disk~\citep{helmi2018merger}. 
In regards to specific effects, it has
been argued that the vertical asymmetries and the Gaia snail
could be an aftermath of a collision of the Sagittarius (Sgr) Dwarf
galaxy with the Galactic disk~\citep{widrow2012galactoseismology,Binney2018originGaiaspiral,blandhawthorn2021snail,Gandhi2022snailacrossscales}
with others noting that the last Sgr impact could not have 
been the only source of perturbation~\citep{Bennett2021snailisnotsgr,Bennett2022Exploring_Sgr-MW}. We note, too, that numerical simulations with vertical
asymmetries~\citep{GConde2022phasespirals} or with bending and
breathing features \citep{blandhawthorn2021snail,Hunt2021resolvinglocal,ghosh2022agedissection, hunt2022multiple} can
give rise to phase spirals, 
suggesting that the different phenomena are intertwined.

The interpretations we have noted 
have been made in a one-body picture.
Thus 
further 
simulation and analysis 
are needed 
to compare more easily with the 2PCF results we have found. 
While we leave the task of explicit simulations to future work, we can nonetheless comment on the 
possible theoretical origins for the structures we find. 
First, as motivated in 
Sec.~\ref{sec:ModelModel} and depicted in Fig.~\ref{fig:asym_2pcf_cf}, the over- and under-densities of the vertical waves \citep{widrow2012galactoseismology, yanny2013stellar, bennett2018vertical} appear to map directly onto our vertical structure (z2PCF) results.  Indeed, the z2PCF analysis brings the wave-like features into sharper focus, particularly at larger 
relative separations $z_{12}$. Additionally, that 
this
vertical structure appears to vary across the disk 
agrees broadly with the findings of \citet{ferguson2017milky} and the corrugations suggested by \citet{blandhawthorn2021snail}.  
While \citet{blandhawthorn2021snail} detail via simulation how the Gaia phase-space spiral \citep{Antoja2018youngperturbedMWdisk} may have 
come about due to a superposition of density waves and bending waves caused by 
a collision with the Milky Way, assessing a full picture of the 
comparison between 
phase-space structures and
a purely spatial 
2PCF requires a detailed study. 
Nevertheless, 
the suggestion of \citet{blandhawthorn2021snail} that the complex superposition of these two distinct waves is 
responsible for the phase-space spiral(s) 
is compatible 
with the complex vertical landscape we find.  
Certainly, 
our 2PCF analysis disfavors purely planar vertical waves.  Indeed, the breathing modes suggested by \citet{ghosh2022agedissection} and \citet{hunt2022multiple}, the vertical waves \citep{widrow2012galactoseismology}, and the bending of the Galaxy
\citep{levine2006vertical, poggio2018galactic, skowron2019three, chen2019intuitive} could all plausibly contribute 
to the vertical structures we find, and they could act in varying combinations. 

In fact, \citet{ghosh2022agedissection} suggest that Galactic breathing modes are excited by spiral structure, and the particular phase of a star's breathing mode motion would create a $z$-$\phi$ coupling.  As our sample is represented by a very small slice of the simulation in Figure 3 of \citet{ghosh2022agedissection}, we expect this $z$-$\phi$ coupling to be small in our context.  
This matches the results in Fig.~\ref{fig:zPanel_LR_North} where vertical structure in adjacent azimuthal bins appears to be nearly identical, but azimuthal bins separated by some distance begin to show differences in vertical structure at high $|z|$.  This also matches the suggestion of \citet{ghosh2022agedissection} that the breathing mode amplitude is directly proportional to the height above the plane, $|z|$, and so the small effect only becomes noticeable in our data at high $|z|$.

Similarly, \citet{widmark2022mapping} find evidence for breathing modes in the solar neighborhood and suggest that the pattern speed of the local spiral arm is slower than the rotation of the solar neighborhood stars.  \citeauthor{widmark2022mapping} also claim that their Gaussian Process fit method indicates that the Galactic warp affects only the stars in the thick disk, while the thin disk is largely unaffected.  While it is unclear how such a configuration might occur given the observed warp in both stars \citep{skowron2019three, poggio2018galactic} and HI gas \citep{kerr1957magellanic, burke1957systematic, levine2006vertical}, this behavior does potentially explain the results in Fig.~\ref{fig:zPanel_LR_North}.  That is, the gradual warping introduces a small $z$-$\phi$ coupling only for the thick disk, and thus only the high-$|z|$ peaks separated by some azimuthal distance exhibit a measurable breaking of axial 
symmetry in their vertical structure.  Nonetheless, 
the {\it Gaia} data used by \citet{widmark2022mapping} may be more difficult to interpret 
because of dust and stellar crowding effects in the mid-plane. 
Suggesting that 
the thick and thin disks respond in a observably different way to 
perturbations may also suggest that they acted in 
the relatively recent past, which is intriguing. 

We emphasize that the structural differences we find appear more marked in the North, and it is
likely that this effect cannot be explained solely by the mechanisms we have noted 
in previous paragraphs.  
For context, we recall that 
\citet{Bennett2022Exploring_Sgr-MW, Bennett2021snailisnotsgr} contend that the Sagittarius impact picture alone cannot reproduce either the Gaia snail or the 
vertical waves in simulations.
Moreover, 
\citet{Gandhi2022snailacrossscales} show that the current passage of the Sagittarius Dwarf has already had its greatest effect while below the disk, and is likely to have 
made 
a much smaller effect when it crosses into the North, due to a larger average galactocentric radius.

Regardless, whether the observed perturbations come from 
a superposition of waves 
\citep{blandhawthorn2021snail} or from 
a Sagittarius impact modulated by the Large Magellanic Cloud \citep{laporte2018influence}, it 
seems increasingly clear that a multitude of effects are likely recorded in the structure of the Milky Way -- 
and this is made all the more likely by the growing census of past mergers 
\citep{Malhan2022GlobalDynamicalAtlas, Lovdal2022Substructurestellarhalo}.  If the Milky Way 
did 
grow hierarchically over cosmic time, as expected in 
the cold dark matter paradigm~\citep{Peebles1993ppc..book.....P}, 
the aggregate of its entire merger history and its 
extremely long relaxation time 
\citep{binney2008GD} 
does lend credence to a multi-effect picture. 
Swathes of stars 
perturbed by past impacts with the disk would then retain some information about these long-ago mergers, resulting in compounding alterations to the Galaxy's structure. 
Presumably, the 2PCF analyses effected within this paper measure the integrated properties of these myriad mergers, along with other effects, with the various probes
of symmetry breaking hinting to 
particular effects.  
Comparing 
future simulation with future 2PCF analyses could yield sharpened constraints 
on the various scenarios, 
as the discriminating ability of the 2PCF surpasses that of stellar number count studies 
thanks to their ${\cal O}(N^2)$ statistics.

\section{Conclusions} \label{sec:Conclusion}

We have introduced 
a new realization of the 2PCF and have 
derived useful functional forms for spherical and slab geometries 
in the steady-state limit 
that show that the 2PCF is vanishingly small 
at the length scales probed in our study. 
This provides a setting for our observational analysis in which, 
by exploiting reflection 
and axial symmetry, 
we have compared the structural differences of various regions of the Galaxy against one another.  Particularly, we have examined the 2PCF 
as a function of the 
separation in $x$, $y$, or
$z$ only, for different selections of 
$R,\phi$, and $z$, fingerprinting the effects of the various perturbations that 
have acted over the Galaxy's history. 
As we have developed in Sec.~\ref{sec:Theory}, 
these observed effects attest to the existence of 
time-dependent perturbations. 
Ultimately, it is clear from this analysis that the stars in our Galaxy are not perfectly uncorrelated as commonly assumed \citep{binney2008GD}.  
Rather, we have discovered that the 
stars are highly-correlated in the vertical direction -- confirming and 
sharpening 
previous discoveries of vertical waves in the Milky Way disk \citep{widrow2012galactoseismology, yanny2013stellar, bennett2018vertical}.  These wave-like, vertical structures exhibit small differences in phase and amplitude across the Galactic disk, 
especially at higher $|z|$, and non-adjacent wedges of data show 
marked azimuthal differences in the waves, also at higher $|z|$.  

In addition to the aforementioned vertical structures, we find 
substantial evidence for radial and azimuthal structures in the 2PCF. 
To summarize: 
\begin{itemize}
\item
Substantial radial structure appears 
at lower $|z|$, though it is 
much more apparent in the Northern hemisphere. 
Radial structure in the North extends all the way  
to 1.2 kpc above the plane, while radial structure in the South is mostly confined to $0.2 < |z| < 0.3$ kpc.  This North-heavy structure trend is 
consistent with the azimuthal differences in vertical structure seen in the North, and with the relative lack of structure seen in the South. 

\item
Further, some hints of azimuthal structure exist -- again predominantly in the North.  
Some slight azimuthal correlations exist at low $|z|$ in the North, 
but otherwise the thin disk appears to be devoid of azimuthal structure at the $z$ 
we consider, which ought be well away from the spiral arms.  
Very interestingly, an azimuthal wave structure with a wavelength around 40 or 50 pc 
is apparent at high $|z|$ in the North, perhaps speaking to previously undiscovered dynamical effects.  

\item Additionally, we find evidence of substantial structural variations across $R$ and $\phi$ in the vertical direction.  Not only have we resolved the vertical waves discovered by \citet{widrow2012galactoseismology} in our z2PCF analysis, a Left-Right comparison shows significant differences at high $|z|$, 
suggesting that the waves are 
being disrupted 
(i.e., not perfectly planar waves, agreeing with the findings of \citet{ferguson2017milky}), or superimposed on an entirely different effect, 
such as 
the corrugations suggested by \citet{blandhawthorn2021snail} or a tilt in the mid-plane location \citep{eilers2020strength, katz2018gaia}.

\end{itemize}

More study is required to determine the precise origin(s) of the correlations we find.
We do think that warping or tilting of the disk is not a sole contributing cause to 
the azimuthal structures that we see in Figs.~\ref{fig:zPanel_LR_North} -~\ref{fig:yPanel_LR_South}. 
Since this analysis is effected in heliocentric coordinates, the true Galactic mid-plane is really below our $z = 0$ plane.  
This in itself would seem 
to disfavor disk warping or tilting
as a sole cause because these effects 
would presumably be visible both in the North and South, which is not at all 
what we observe.

\acknowledgements{
S.G. and A.H. acknowledge partial support from the 
University Research Professor (S.G.) fund of the University of
Kentucky and from the U.S.
Department of Energy under contract DE-FG02-96ER40989. 
A.H. acknowledges support from the Universities Research Association and the University of Kentucky College of Arts \& Sciences' Dean's Competitive Fellowship.  

We thank Joss Bland-Hawthorn for early remarks that we considered in framing our systematic error analysis and Scott Tremaine for comments on the accepted paper that we have addressed in the proofs.
We also thank the anonymous referee for helpful comments that have improved the presentation of our paper.  

This document was prepared in part using the resources of Fermi National Accelerator Laboratory (Fermilab), a U.S. Department of Energy, Office of Science, HEP User Facility. Fermilab is managed by the Fermi Research Alliance, LLC (FRA), acting under Contract No. DE-AC02-07CH11359. 

This work has made use of data from the European Space Agency (ESA) mission
{\it Gaia} (\url{https://www.cosmos.esa.int/gaia}), processed by the {\it Gaia}
Data Processing and Analysis Consortium (DPAC,
\url{https://www.cosmos.esa.int/web/gaia/dpac/consortium}). Funding for the DPAC
has been provided by national institutions, in particular the institutions
participating in the {\it Gaia} Multilateral Agreement.
}

\bibliography{editable_biblio_big.bib, mybib.bib}

\end{document}